\documentclass[12pt]{article}

\usepackage{color}
\usepackage[english]{babel}
\usepackage{amssymb,amsmath}
\usepackage{graphicx}
\usepackage[colorlinks=true,linkcolor=blue]{hyperref}
\usepackage{cleveref}
\usepackage{physics}
\usepackage{fontawesome5}
\usepackage[dvipsnames]{xcolor}
\usepackage{color}
\usepackage{upgreek}
\usepackage{caption}
\usepackage{feynmp}
\usepackage{tikz}
\newcommand{\github}[1]{%
   \href{#1}{\faGithubSquare}%
}
\def\ba{\begin{eqnarray}}
\def\ea{\end{eqnarray}}
\def\be{\begin{equation}}
\def\ee{\end{equation}}

\def\d{\partial}

\renewcommand{\l}{\lambda}
\renewcommand{\O}{\Omega}

\newcommand{\g}{\gamma}

\newcommand{\diff}{\mathrm{d}}
\newcommand{\e}{{\rm e}}

\renewcommand{\a}{\alpha}

\renewcommand{\g}{\gamma}
\newcommand{\s}{\sigma}

\newcommand{\ve}{\varepsilon}
\newcommand{\T}{\mathcal{T}}

\newcommand\Teff{\mathcal{T}_{\rm eff}}

\newcommand{\PP}{\mathcal{P}}

\newcommand\bseq{\begin{subequations}}
\newcommand\eseq{\end{subequations}}

\usepackage{etoolbox}
\preto\subequations{\ifhmode\unskip\fi}

\allowdisplaybreaks[4]

\begin{document}

\title{
Simple third order operator-splitting schemes \\for stochastic mechanics and field theory
}

\author{
\\
Andrey Shkerin$^{1}$\thanks{ashkerin@perimeterinstitute.ca}~,~ 
Sergey Sibiryakov$^{1,2}$\thanks{ssibiryakov@perimeterinstitute.ca}
\\[2mm]
{\small\it $^1$Perimeter Institute for Theoretical Physics,}
{\small\it Waterloo, ON N2L 2Y5, Canada,}\\
{\small\it $^2$Department of Physics \& Astronomy, McMaster University,}
{\small\it Hamilton, ON L8S 4M1, Canada}
}

\date{}

\maketitle

\begin{abstract}
We present a method for constructing numerical schemes with up to 3rd strong convergence order for solution of a class of stochastic differential equations, including equations of the Langevin type. The construction proceeds in two stages. In the first stage one approximates the stochastic equation by a differential equation with smooth coefficients randomly sampled at each time step. In the second stage the resulting regular equation is solved with the conventional operator-splitting techniques. This separation renders the approach flexible, allowing one to freely combine the numerical techniques most suitable to the problem at hand. The approach applies to ordinary and partial stochastic differential equations. In the latter case, it naturally gives rise to pseudo-spectral algorithms. We numerically test the strong convergence of several schemes obtained with this method in mechanical examples. Application to partial differential equations is illustrated by real-time simulations of a scalar field with quartic self-interaction coupled to a heat bath. The simulations accurately reproduce the thermodynamic properties of the field and are used to explore dynamics of thermal false vacuum decay in the case of negative quartic coupling.\footnote{
The codes used in this work are available on GitHub: \url{https://github.com/Olborium/Stoch_Integ}.}

\end{abstract}

\newpage
{\hypersetup{hidelinks}
\tableofcontents
}

\section{Introduction}
\label{sec:intro}

Operator-splitting methods for numerical solution of ordinary and partial differential equations provide an important toolkit for the analysis of nonlinear dynamics \cite{McLachlan_Quispel_2002,Blanes_Casas_Murua_2024}. In the case of Hamiltonian systems they lead to symplectic integrators which preserve exactly the canonical structure of the evolution and thereby greatly enhance the accuracy of the numerical scheme \cite{mclachlan1992accuracy,hairer2013geometric}. For partial differential equations, they form the basis of pseudo-spectral algorithms, with part of the evolution performed in configuration space and part in the Fourier space. This typically results in superior stability and accuracy of the code compared to the traditional finite difference schemes \cite{mclachlan1993explicit}. There is a general technique for constructing operator-splitting schemes of arbitrarily high convergence order for deterministic evolution equations \cite{yoshida1990construction} 
(see \cite{Blanes_Casas_Murua_2024} and references therein for methods of up to 8th order)
and resulting algorithms are widely used in many branches of research \cite{duffy2013finite,leimkuhler2015molecular,Bou_Rabee_2018}, including field theory and cosmology \cite{Felder:2000hq,Frolov:2008hy,Huang:2011gf,Sainio:2012mw,Amin:2018xfe,Lozanov:2019jff,Figueroa:2020rrl}.

This is in stark contrast to the case of stochastic differential equations (SDEs) which contain forcing terms driven by Gaussian white noise. Such equations are ubiquitous in mathematical modeling across multiple fields of science. They appear in finance, data analysis, biology, chemistry, and physics \cite{mao2007stochastic,allen2007modeling,duffy2013finite,oksendal2013stochastic,sobczyk2013stochastic,leimkuhler2015molecular}. 
Yet the numerical algorithms for them rarely go above the convergence order\footnote{We talk here about the strong convergence order which directly characterizes the difference between the numerical and exact solutions, as opposed to the weak order referring to their statistical properties \cite{kloeden2013numerical}.} $1.5$. This is due to the singular nature of the white noise which makes impossible a straightforward adaptation of the methods developed for deterministic equations and relying on smoothness of the equation coefficients and the solution. Application of a proper generalization of the Taylor expansion (It\^o--Taylor or Stratonovich--Taylor) shows that stochastic numerical schemes involve iterated time integrals of the white noise \cite{kloeden2013numerical}. For a general SDE, quadratic integrals of the noise arise at orders higher than $1.5$, which obey complicated non-Gaussian statistics. This makes their numerical sampling impractical. 

Fortunately, it is known \cite{hershkovitz1998fourth,milstein2003quasi,milstein2021stochastic} 
that the appearance of non-Gaussian terms is postponed till the 4th order for an important subclass of SDEs known as (underdamped) Langevin equations with additive noise. Equations of this type play major role in all above research areas \cite{risken1996fokker,snook2006langevin,coffey2012langevin,leimkuhler2015molecular} and have the general form,
\be
\label{KG}
\ddot q+\eta \dot q -f(q)= \sum_r \sigma_r \xi^r\;,
\ee
where $q$ is a $d$-dimensional coordinate vector, dot stands for time derivative, $\eta$ is the matrix of friction coefficients, $f(q)$ is the coordinate dependent force, $\sigma_r$ are constant vectors, and $\xi^r(t)$ are independent normalized Gaussian white noises
\be
\label{Noise0}
\langle\xi^r(t)\rangle=0\;,\qquad\langle \xi^r(t)\xi^s(t')\rangle=\delta^{rs} \, \delta(t-t')\;.
\ee
The classical occurrence of \eqref{KG} in physics is in the context of Brownian motion of a particle interacting with environment. The force in this case can often be expressed as the gradient of potential energy,
\be
\label{potforce}
f(q)=-\nabla U(q)\;,
\ee
and, if the environment is in thermal equilibrium, the noise and friction are related by the fluctuation-dissipation theorem \cite{Landau:1980mil}. 

Equation (\ref{KG}) can be generalized to field theory. An example is the dissipative nonlinear Klein--Gordon equation describing a scalar field $\phi(t,x)$ in (1+1)-dimensional spacetime interacting with a heat bath \cite{Pirvu:2024ova,Pirvu:2024nbe},
\be \label{LangEq}
\ddot\phi + \eta\dot\phi - \phi^{\prime\prime} + \phi + \a \phi^3 = \sqrt{2\eta \T} \:\xi \;,
\ee 
where prime denotes the spatial derivative, $\T$ is the temperature of the bath, 
and the noise is uncorrelated both in time and space, 
\be 
\label{Noise1}
\ev{\xi(t,x)} = 0 \;, \qquad \ev{\xi(t,x) \xi(t^{\prime},x^{\prime})} = 
\delta(t-t^{\prime}) \delta(x-x^{\prime}) \;.
\ee
The parameter $\a$ takes values $+1$ or $-1$ corresponding to repulsive or attractive self-inte\-rac\-tion, respectively.

There exists a number of numerical schemes for eq.~(\ref{KG}) involving only Gaussian random variables, including operator-splitting schemes \cite{10.1093/imanum/dru056,buckwar2020spectral,milstein2021stochastic,monmarche2021high,CuiHongSheng,song2023hamiltonian}. The latter exist up the the 3rd order \cite{hershkovitz1998fourth,milstein2003quasi,milstein2021stochastic,Foster2021}, but they
are rather sophisticated and not easily adaptable for different choices of splitting which may be beneficial in various applications. We are not aware of any uses of these schemes in field theory. In particular, the standard numerical packages \cite{dennis2013xmds2,kiesewetter2016xspde} for solution of stochastic partial differential equations typically provide 1st strong order convergence and do not include any schemes going beyond 2nd strong stochastic order. 

The purpose of this paper is to present a simple and flexible 3rd order operator-splitting method for a class of SDEs with additive noise, containing eqs.~(\ref{KG}) and (\ref{LangEq}) as subcases. Our main idea follows Refs.~\cite{castell1996ordinary,Misawa2000,misawa2001lie,foster2020optimal,telatovich2017} and consists of two steps. In the first step, the time interval $0\leqslant t\leqslant T$ is split into $\mathcal{N}$ sub-intervals of length $h$. Within each sub-interval $t_n\leqslant t \leqslant t_{n+1}$, $n=1,..,\mathcal{N}$, the SDE is replaced by an ordinary differential equation (ODE) with randomly drawn coefficients. The coefficients are time-independent within sub-intervals, but are different for different sub-intervals. They are chosen such that the mean absolute difference between the solutions of the original SDE and the ODE is $\mathcal{O}(h^3)$ at the endpoints of the time steps $t=t_n$, $n=1,..,\mathcal{N}$.
In the second step, the approximating ODE is solved on the sub-intervals $t_n\leqslant t \leqslant t_{n+1}$ using any of the standard operator splitting schemes which are most suitable for the problem at hand.
As we are going to show, the outlined method is easily adapted to field theory.


The paper is organized as follows. In Sec.~\ref{sec:theory} we describe the approximation of a class of SDEs by ODEs, accessible to standard operator-splitting methods. We argue that the approximation possesses 3rd strong convergence order and provide examples of its application to Langevin equation (\ref{KG}) and its field theory generalization (\ref{LangEq}).
In Sec.~\ref{sec:particle} we numerically test the strong convergence of the scheme on simple mechanical systems and demonstrate its superiority with respect to lower-order algorithms. We illustrate the flexibility of the method by combining the stochastic approximation with different 
operator splittings. 

In Sec.~\ref{sec:field} we promote the scheme to a pseudo-spectral algorithm and apply it to the field theory. We consider eq.~(\ref{LangEq}) with $\a=1$ and evolve the field towards thermal equilibrium. We analyze the weak convergence of the key thermodynamic observables, such as the effective temperature and power spectrum of the field modes. We then consider the case $\a=-1$ which corresponds to a field theory with unstable quartic potential and measure the decay rate of the metastable state $\Gamma$ as function of the dissipation coefficient, exploring a range from extremely weak ($\eta\sim 10^{-3}$) to strong ($\eta\sim 50$) damping. In the overdamped regime $\eta\geq 10$ we reproduce the classic result $\Gamma\propto \eta^{-1}$ \cite{Kramers:1940zz,Langer:1969bc}, thereby complementing the study of Refs.~\cite{Pirvu:2024ova,Pirvu:2024nbe}.

We conclude in Sec.~\ref{sec:disc}. Appendices contain derivation of the approximation scheme and other technical details. 

The codes used in this paper are written in C++ 17 (numerical schemes, solution generation) and Python 3 (processing routines, making plots). They are available on GitHub \cite{GitHub}. We ran the C++ codes on the general-purpose clusters Fir and Narval of the Digital Research Alliance of Canada \cite{Alliancecan}, using CPU nodes with 2 $\times$ AMD EPYC 9655 (Zen 5) \@ 2.7 GHz processors. The Python codes were run on the Symmetry cluster of the Perimeter Institute \cite{Symmetry}, using head nodes with Intel Xeon Silver cores.


\section{Approximating stochastic equation with an ODE}
\label{sec:theory}

\subsection{General scheme}
\label{ssec:scheme}

We consider a system of SDEs,\footnote{We use the convention of summing over repeated indices, unless stated otherwise.}
\bseq
\label{NLstoch+}
\begin{align}
\label{NLstoch}
&\dot z^i=F^i(z)+\s^i_{\;r}\,\xi^r\;,\\
&z^i\big|_{t=0}=z_0^i\;,~~~~~i=1,\ldots,I\;,~~~~r=1,\ldots,R\;.
\end{align}
\eseq
$R$-dimensional Gaussian white noise $\xi^r(t)$ 
satisfies eqs.~(\ref{Noise0}), whereas the $I\times R$ matrix 
$\s^i_{\;r}$ is constant, i.e. the noise in eq.~(\ref{NLstoch}) is additive.  
The deterministic force $F^i(z)$ is assumed to be sufficiently smooth: the analysis of the 3rd order strong convergence requires continuous derivatives of $F^i(z)$ up to the 4th order; see Appendix~\ref{ssec:der} for details.
We also assume that the columns of the matrix $\s^i_{\;r}$ lie in the null space of the Hessian of the deterministic force $F^i(z)$,
\be
\label{Hess}
F^i_{\;,jk}\,\s^j_{\;r}=0\;.
\ee
Here we use coma to denote derivatives with respect to the coordinates, $F^i_{\;,j}=\frac{\d F^i}{\d z^j}, F^i_{\;,jk}=\frac{\d^2F^i}{\d z^j \d z^k}$, etc.
The condition (\ref{Hess}) is important: as detailed in Appendix~\ref{ssec:der}, it ensures that \cref{NLstoch} can be approximated up to the 3rd strong order using Gaussian random variables only. 
As we are going to see shortly, it is naturally satisfied by the Langevin-type \cref{KG,LangEq} but, in principle, is more general.

Under these assumptions the solution of (\ref{NLstoch+}) can be approximated within a small time interval $0\leq t\leq h$ by the solution of an ODE, 
\bseq
\label{Langevin2+}
\begin{align}
\label{Langevin2}
&\dot {\hat z}^i=F^i(\hat z)+\s^i_{\;r} \zeta_1^r+F^i_{\;,j}(\hat z)\s^j_{\;r}\zeta_2^r
+F^i_{\;,j}(\hat z)F^j_{\;,k}(\hat z)
\s^k_{\;r}\zeta_3^r\;,
\\
&\hat z^i\big|_{t=0}=z^i_0\;,
\end{align}
\eseq
where $\zeta^r_a$, $a=1,2,3$, are {\it time-independent} random variables related to the original white noise as  
\bseq\label{newrandvars}
\begin{align}
\label{newrandvar1}
&\zeta_1^r=\frac{1}{h}\int_0^h \diff \tau\, \xi^r(\tau)\;,\\
\label{newrandvar2}
&\zeta_2^r=\frac{1}{h}\int_0^h \diff \tau\, \bigg(\frac{h}{2}-\tau\bigg)
\xi^r(\tau)\;,\\
\label{newrandvar3}
&\zeta_3^r=\frac{1}{h}\int_0^h \diff \tau\, 
\bigg(\frac{h^2}{12}-\frac{h\tau}{2}+\frac{\tau^2}{2}\bigg)
\xi^r(\tau)\;.
\end{align}
\eseq
In Appendix~\ref{ssec:der} we show that the difference between the solutions of eqs.~(\ref{NLstoch+}) and (\ref{Langevin2+}) obeys the properties,
\be
\label{diff}
\big\vert \big\langle z(h)-\hat z(h)  \big\rangle\big\vert=\mathcal{O}(h^4)~,~~~~
\big\langle \big(z(h)-\hat z(h)\big)^2\big\rangle=\mathcal{O}(h^7)\;.
\ee
Here, in the first expression the average over the stochastic ensemble is taken first, and then the absolute value of the average is evaluated. Whereas, in the second expression, we first square the difference between the exact and approximate solutions, and then average over the ensemble.
Note that we do not claim a uniform approximation of the true solution of the SDE (\ref{NLstoch}) $z(t)$ by the ODE solution $\hat{z}(t)$ inside the interval $0\leqslant t\leqslant h$. Indeed, such an approximation would be impossible due to very different regularity properties of the two functions. Equations (\ref{diff}) only imply that the two solutions are close to each other at the \textit{end-points} of the small time interval. This property is sufficient for our purposes.

Imagine now that we have split a finite interval $0\leq t\leq T$ into ${\cal N}$ steps of length $h$. Within each step $t_n\leq t\leq t_{n+1}$, $t_n=nh$, we define an ODE of the form (\ref{Langevin2}), with its own value of the parameters $\zeta_{a|n}^r$. We then solve these equations consecutively, taking for the initial data at the $n$th step the final value 
$\hat z(t_n)$ from the previous step. {\it Proposition 1} from Appendix~\ref{ssec:der} 
shows that the resulting function differs from the true solution of the stochastic equation (\ref{NLstoch+}) at the grid points by no more than $\mathcal{O}(h^3)$,
\be
\label{major3}
\langle |z(t_n)-\hat z(t_n)|\rangle \leq C\,h^3\;, ~~~~~n=0,1,\ldots,{\cal N}\;,
\ee
for some constant $C$ which does not depend on ${\cal N}$. In other words, if we solve the ODEs (\ref{Langevin2}) in each time step numerically with sufficient accuracy, we obtain an approximation converging to $z(t)$ at the 3rd strong order.\footnote{Note that, unlike the first of eqs.~(\ref{diff}), here the absolute value of the difference between the true and approximate solutions is taken \textit{before} averaging.}

In practice, we do not need to compute the parameters $\zeta^r_a$ from the underlying white noise using eqs.~(\ref{newrandvars}).\footnote{This is required, however, for tests of strong convergence which compare numerical approximations to $z(t)$ for different sizes of the time step $h$ at a fixed realization of the noise, see Sec.~\ref{sec:particle}.} We only need to reproduce their statistical properties. Clearly, they are Gaussian random variables with zero mean. Further, they happen to be uncorrelated:
\bseq
\label{Cab}
\begin{align}
&\langle \zeta_a^r \,\zeta_b^s\rangle=\mathcal{C}_a\,\delta_{ab}\,\delta^{rs}~~~~~\text{(no summation over $a$)}\\
\label{Ca}
&\mathcal{C}_{1}=\frac{1}{h}~,~~~~~
\mathcal{C}_{2}=\frac{h}{12}~,~~~~~
\mathcal{C}_{3}=\frac{h^3}{720}\;.
\end{align}
\eseq
The values of $\zeta_a^r$ at different time steps are also independent, which makes their numerical sampling straightforward. 

Once $\zeta_a^r$ have been drawn, eq.~(\ref{Langevin2}) within a given time step can be integrated using standard numerical methods for ODEs, including operator-splitting schemes of arbitrarily high order. Of course, choosing the order of splitting significantly higher than 3 is not expected to give any practical advantages since the accuracy of the algorithm will be set by the difference between the approximating ODE (\ref{Langevin2}) and the original SDE (\ref{NLstoch}). On the other hand, the splitting order should not be lower than 3, to fully exploit the power of the stochastic approximation. Below we will illustrate the use of several splitting methods with explicit examples. 

Let us summarize our algorithm to obtain a numerical solution to SDE (\ref{NLstoch+}) of 3rd strong convergence order:
\begin{enumerate}
\item 
Discretize the time interval in steps of length $h$.
\item 
For each time step $t_n\leq t\leq t_{n+1}$, $t_n=nh$, generate a triplet of Gaussian random variables $\zeta_{a|n}^r$ with diagonal correlation matrix (\ref{Cab}). The variables in different time steps are independent. 
\item
Inside each time step replace the original SDE with the ODE (\ref{Langevin2}).
\item 
Solve the resulting sequence of ODEs using any method with the convergence order higher or equal to 3. This can be done using the standard operator-splitting techniques for ODEs.
\end{enumerate}
In Appendix~\ref{app:algorithm} we formally present two numerical schemes based on this algorithm to solve the Langevin equation (written in the first-order form, see \cref{Langevin0} below).
Two comments are in order:
\begin{itemize}
\item[i)]
Without loss of precision, the coordinate $\hat z$ in the term containing $\zeta_3$ in eq.~(\ref{Langevin2}) at the $n$th time step can be replaced by its value $\hat z(t_n)$ in the beginning of the step. This renders the $\zeta_3$-term constant over the entire time step.
Due to the condition (\ref{Hess}), the term with $\zeta_2$ is actually also constant, so that the whole random force in (\ref{Langevin2}) can be evaluated only once in the beginning of the time step. This considerably accelerates the numerical solution of the equation.  
\item[ii)] 
Truncation of eq.~(\ref{Langevin2}) discarding the $\zeta_3$-term produces a scheme of 2nd strong order which is equivalent to the log-ODE method of Ref.~\cite{foster2020optimal} restricted to the case of additive noise with the property (\ref{Hess}). In fact, the analysis of Appendix~\ref{ssec:der} shows that this scheme provides 2nd strong convergence under a weaker assumption,
\be
\label{Hess1}
F^i_{\;,jk}\,\s^j_{\;r}\,\s^k_{\;s}=0\;.
\ee
Discarding both terms with $\zeta_3$ and $\zeta_2$ leads to the well-known 1st order Euler--Ma\-ru\-ya\-ma method. 
\end{itemize}
We now illustrate the above scheme with a few examples.

\subsection{Langevin equation}
\label{ssec:LangEq}

We start with the Langevin eq.~(\ref{KG}) which we rewrite in the first-order form,
\be\label{Langevin0}
\left\lbrace \begin{array}{l}
   \dot p^i=-\eta p^i+f^i(q)+\s^i_{\;r} \,\xi^r\;\\
    \dot q^i=p^i
\end{array}\right.,
\qquad i=1,\ldots,d~,~~~~~r=1,\ldots,R\;.
\ee
For simplicity, we have assumed here that the matrix of the dissipative coefficients is proportional to a unit matrix.
Combining the phase space coordinates into a single vector with components
\be
z^i=\begin{cases}
p^i~,& 1\leq i\leq d\;\\
q^{i-d}~,& d<i\leq 2d\;
\end{cases}
\ee
we obtain eq.~(\ref{Langevin2}) with 
\be
\label{Langforce}
F^i=\begin{cases}
-\eta p^i+f^i(q)~,& 1\leq i\leq d\;\\
p^{i-d}~,& d<i\leq 2d\;
\end{cases}
\ee
and the noise matrix $\s^i_{\;r}$ with non-vanishing components only at $i\leq d$. 
Since the force $F^i$ is linear in momenta, its second derivatives satisfy the condition (\ref{Hess}).  The function $f^i(q)$ must be sufficiently smooth; see the discussion below \cref{NLstoch+}. In the concrete examples below, $f^i(q)$ will be an infinitely continuous differentiable function of the coordinates.
The approximating ODE 
at the $n$th time step $t_n\leq t\leq t_{n+1}$ takes the form,\footnote{Equations (\ref{Langevin1}) appear in Ref.~\cite{telatovich2017}. However, their correlation matrix for the variables 
$\zeta^r_a$ differs from (\ref{Cab}). We believe this discrepancy is due to a mistake in the earlier work \cite{Kunita1980}, whose results are used in \cite{telatovich2017}.}
\be \label{Langevin1}
\left\lbrace \begin{array}{l}
    \dot p^i = -\eta p^i +f^i(q) + \s^i_{\;r} \,\zeta_{1|n}^r 
    - \eta \s^i_{\;r} \,\zeta_{2|n}^r
    + \big(\eta^2\delta^i_j +f^i_{\;,j}\big(q_n\big)\big)\s^j_{\;r} \,\zeta_{3|n}^r \;\\
    \dot q^i = p^i + \s^i_{\;r}\, \zeta_{2|n}^r - \eta \s^i_{\;r}\,\zeta_{3|n}^r \;
\end{array} \right.
\ee
Note that in the last term of the first equation we used the simplification mentioned in the point (i) above, namely we substituted $q$ by its value in the beginning of the step $q_n\equiv q(t_n)$. In this way all the terms representing the stochasticity become constant within the time step.

The solution to eq.~(\ref{Langevin1}) 
on a single time step is found 
using standard operator-splitting methods. Let $\big(p(t_n),q(t_n)\big)$ be the solution at time $t_n$, then at time $t_{n+1}=t_n+h$ the solution can be written as
\be \label{EqSol1}
\left( \begin{array}{l} p(t_{n+1}) \\ q(t_{n+1}) \end{array} \right) = \e^{h\hat{O}} \left( \begin{array}{l} p(t_{n}) \\ q(t_{n}) \end{array} \right) \;.
\ee
The idea of the splitting technique is to replace the evolution operator $\e^{h\hat{O}}$ with the product of operators $\e^{h\hat{A}}$, $\e^{h\hat{B}}$ such that $\hat{O}=\hat{A}+\hat{B}$ and the evolution due to $\hat{A}$ and $\hat{B}$ can be computed exactly. 
Two types of splitting are commonly used.
The first (PQ splitting) separates the evolution of the coordinates from that of the momenta. Denote\footnote{To avoid clutter, we omit the index corresponding to the number of the time step.}
\be \label{Mu12}
\mu_1^i =\s^i_{\;r} (\zeta_2^r - \eta\zeta_3^r) \;, ~~~ 
\mu_2^i = \s^i_{\;r}\big(\zeta_1^r - \eta\zeta_2^r + \eta^2\zeta_3^r)
+f^i_{\;,j}\big(q_n\big)\s^j_{\; r}\,\zeta_3^r \;.
\ee 
The operators ${\hat A}$, ${\hat B}$ correspond to the following equations:
\be \label{PQ}
\begin{aligned}
& \hat{A}:~~~~ \dot{p}^i = 0 \;, ~~~\dot{q}^i = p^i+\mu_1^i\;, \\
& \hat{B}:~~~~ \dot{p}^i = -\eta p^i +f^i(q) +\mu_2^i \;, ~~~ \dot{q}^i = 0 \;. 
\end{aligned}
\ee
Clearly, each of these pairs of equations can be solved exactly on $(t_n,t_{n+1})$.
The second type (LN splitting) separates the evolution of the linear and nonlinear parts of the system.
Let us split the force as $f=f_{\rm lin}+f_{\rm int}$ where $f_{\rm lin}(q)$ is defined to be at most linear in $q$, while $f_{\rm int}(q)$ contains all nonlinearities. Then the operators ${\hat A}$, ${\hat B}$ are associated with the equations
\be \label{LN}
\begin{aligned}
& \hat{A}:~~~~ \dot{p}^i = -\eta p^i +f^i_{\rm lin}(q) 
+\mu_2^i \;, ~~~ \dot{q}^i = p^i + \mu_1^i\;, \\
& \hat{B}:~~~~ \dot{p}^i = f^i_{\rm int}(q)\;, ~~~~\dot{q}^i = 0 \;. 
\end{aligned}
\ee
Again, these equations admit an exact solution. Note that the replacement $q(t)\mapsto q(t_n)$ in the stochastic term $\mu_2^i$ has allowed us to include it into the linear evolution. Without this replacement, the noise would contribute into the nonlinear part as well. 

To compute the solution at $t=t_{n+1}$, in \cref{EqSol1} one replaces
\be \label{Split}
\e^{h{\hat O}} = \prod_{l=s}^1 (\e^{h a_l {\hat A}}\e^{h b_l {\hat B}}) \;,
\ee
where $a_l$, $b_l$ are splitting coefficients found by minimizing the error between the exact and approximate evolution operators. The parameter $s$ is the order of splitting, and in the case of regular differential equations it typically coincides with the order of convergence.
The coefficients for the splitting orders 3 and 4 
can be found, e.g., in \cite{mclachlan1992accuracy}. In Sec.~\ref{sec:particle} we numerically solve eq.~(\ref{Langevin1}) using the PQ and LN splittings of different orders and 
test the strong convergence of the algorithm in the case of stochastic mechanics with one degree of freedom. In Appendix~\ref{app:algorithm} we give explicit examples of algorithms implementing the 3rd strong order scheme with 4th order PQ and LN splittings for \cref{Langevin0}.

The above construction admits a straightforward generalization to the case when the force 
in (\ref{Langevin0}) depends linearly on momenta with arbitrary constant coefficients, like e.g. for a charged particle moving in a constant magnetic field. If the magnetic field is position dependent, the property (\ref{Hess}) will in general be violated and the 3rd strong order scheme does not apply. Still, the weaker property (\ref{Hess1}) will be satisfied even in this case, so one can use the 2nd order truncated scheme obtained by discarding the $\zeta_3$ terms in (\ref{Langevin1}) (see comment (ii) in Sec.~\ref{ssec:scheme}).

\subsection{Pseudo-spectral algorithm for field theory}
\label{ssec:fieldtheory}

The LN splitting is particularly useful when solving partial differential equations, since the linear evolution can be carried out in the Fourier space. The transformation between the real and Fourier spaces is performed using the Fast Fourier Transform. Numerical schemes implementing this technique are referred to as pseudo-spectral schemes. 

We illustrate this point using eq.~(\ref{LangEq}). We discretize the spatial variable $x$ 
on a lattice of length $L$, with $N$ sites and periodic boundary conditions. The lattice sites are labeled by ${x_i\equiv \ell  \, i}$, where $\ell=L/N$ is the lattice spacing, and $i=0, \dots ,N-1$; we set ${x_N\equiv x_0}$. The physical degrees of freedom are $\phi^i\equiv\phi(x_i)$. Using second-order finite-difference approximation for the spatial Laplacian and introducing the canonically conjugate momenta $\pi^i=\dot\phi^i$
we obtain a system of equations,
\be \label{eom1}
\left\lbrace \begin{array}{l}
   \dot \pi^i=-\eta \pi^i +(\Delta \phi)^i - \phi^i - \a (\phi^i)^3  +\s \xi^i\\
   \dot \phi^i=\pi^i  
\end{array}\right., ~~~~~i=0,\ldots,N-1\;,
\ee
with $(\Delta \phi)^i=\ell^{-2}(\phi^{i+1}-2\phi^i+\phi^{i-1})$ and $\s=\sqrt{2\eta\T/\ell}$. Here $\xi^i(t)$ are independent normalized Gaussian white noises.
Note that we do not attempt to improve the spatial discretization of the noise. This is consistent with interpreting the lattice theory as a Hamiltonian system coupled to thermal bath \cite{Pirvu:2024nbe}.

We observe that the system (\ref{eom1}) has the form (\ref{Langevin0}). Thus we can immediately write down the approximating ODE for it inside the $n$th time step $t_n\leq t\leq t_{n+1}$, 
\be \label{Langevin3}
\left\lbrace \begin{array}{l}
   \dot \pi^i=-\eta \pi^i +(\Delta \phi)^i\! -\! \phi^i\! -\! 
   \a (\phi^i)^3  
   +\s \zeta_{1|n}^i \!-\!\eta\s\zeta_{2|n}^i 
   + \eta^2\s \zeta_{3|n}^i 
   +\s(\Delta \zeta_{3|n})^i \!-\! 3\a\s \big(\phi^i_n\big)^2 \zeta_{3|n}^i\\
    \dot \phi^i=\pi^i + \s \zeta_{2|n}^i - \eta\s \zeta_{3|n}^i   
\end{array}\right.
\ee
where $\phi^i_n\equiv \phi^i(t_n)$.
The discrete noise variables have the diagonal correlation matrix (\ref{Cab}). 
This approximation has the 3rd order strong convergence in the size of the time step $h$.

To find the solution, we split the evolution operator in eq.~(\ref{Langevin3}) into the linear ($\hat{A}$) and nonlinear ($\hat{B}$) parts as per \cref{LN}.
The linear part is conveniently solved in the Fourier space where the lattice Laplacian reduces to a simple multiplication.
We expand the field variables into discrete Fourier harmonics
\be \label{IFT}
\phi^i = \frac{1}{\sqrt{N}}\sum_{j=0}^{N-1}\e^{ik_jx_i}\tilde\phi^j \;, 
\qquad k_j=\frac{2\pi j}{L} \;,
\ee
and similarly for $\pi^i$. The condition that the variables in coordinate space are real imposes the relations,
\be
\label{real}
\tilde\phi_j^* = \tilde\phi_{N-j}\;,\qquad \tilde\pi_j^* = \tilde\pi_{N-j}\;.
\ee
Under action of $\hat A$ different modes evolve independently and satisfy 
the equation for a damped harmonic oscillator in the presence of regular external forces:
\be \label{EqsL2}
\hat{A}: \quad
\dot{\tilde\pi}^j = - \eta \tilde\pi^j-(\O^j)^2 \tilde\phi^j  + \tilde\mu^j_2 \;,\qquad
\dot{\tilde\phi}^j = \tilde\pi^j + \tilde\mu^j_1 \;, \\
\ee
where $\O^j$ are the lattice mode frequencies
\be \label{Oj}
(\O^j)^2 = \frac{2}{\ell^2}\big(1-\cos (\ell k_j)\big) + 1 \;.
\ee
The forces are given by
\be \label{nu12}
\tilde\mu_1^j=\s(\tilde\zeta_2^j - \eta\tilde\zeta_3^j) \;, 
\qquad \tilde\mu_2^j =\s\big(\tilde \zeta_1^j - \eta\tilde\zeta_2^j+
\big(\eta^2-(\O^j)^2\big) )\tilde\zeta_3^j-3\s\a \tilde\Xi_3^j\;,
\ee
where $\tilde\zeta_a^j$ are the discrete Fourier images of the variables $\zeta_a^i$, and $\tilde \Xi_3^j$ is the Fourier transform of the combination $\Xi_3^i=\big(\phi^i(t_n)\big)^2\zeta_3^i$.
Explicit solution to eqs.~(\ref{EqsL2}) for a finite fractional time step $ha$
is given in Appendix~\ref{app:pseudo}.

By contrast, the nonlinear part $\hat B$ is evolved in the coordinate space where it corresponds to a local map. It is straightforward to write it for an arbitrary finite time interval,
\be \label{SolN2}
\e^{hb\hat{B}}: \quad \phi^i(t+hb) =\phi^i(t)\;, 
\qquad \pi^i(t+hb) = \pi^i(t)-hb\,\a\big(\phi^i(t)\big)^3  \;.
\ee
In Sec.~\ref{sec:field} we implement the above algorithm numerically using 4th order splitting which requires switching between operators $\hat A$ and $\hat B$, and hence between the Fourier and coordinate spaces, eight times per a single time step. This is done using Fast Fourier Transform. Since we use the scheme in which noise variables are constant over the entire time step, we do not need to transform them at every switch, which accelerates the computation. Further, the noise can be directly generated in the Fourier space. We first generate independent Gaussian random variables $\tilde\zeta_a^j$, such that   
\be \label{Noise4}
\langle{\tilde\zeta^j_a \,\tilde\zeta^{*k}_b} \rangle= {\cal C}_{a}\,\delta_{ab}\,\delta^{jk}\;~~~~~\text{(no summation over $a$)},
\ee
with ${\cal C}_a$ given by (\ref{Ca}). 
We then compute $\tilde\Xi_3^j$ by Fourier transforming $\tilde\zeta_3^j$ to the coordinate space, multiplying by $\big(\phi^i(t_n)\big)^2$, and Fourier transforming back. 
This must be done only once in the beginning of each time step.

\section{Numerical tests of strong convergence}
\label{sec:particle}

In this section we numerically test the strong convergence of the stochastic schemes described above. Since the tests require storing complete realizations of the white noise on a fine time grid, they are memory-heavy. Thus, we restrict them to two simple mechanical systems: stochastic pendulum and stochastic anharmonic oscillator. We illustrate the flexibility of the approach by combining the approximation of SDE by an ODE 
with different operator-splitting methods for solving the resulting ODE.

\subsection{Stochastic pendulum}
\label{ssec:pend}

As the first example, we consider a pendulum in the potential $U(q)=1-\cos q$, coupled to a heat bath which provides dissipation and noise. Motion of the pendulum is described by eq.~(\ref{Langevin0}) with $d=R=1$, 
\be \label{Pot_pend}
f(q)=-\sin q \;,
\ee 
and $\s=\sqrt{2\eta \T}$. We fix the temperature of the bath to be $\T=1$ and the friction coefficient $\eta=10$. We perform our tests in the overdampted regime to make them more stringent. Indeed, the contributions with the stochastic variables in the approximating ODE (\ref{Langevin1}) get enhanced with increasing $\eta$, the 3rd order term having the strongest scaling $\propto \eta^{5/2}$. Choosing $\eta$ to be large allows us to check that this does not lead to any numerical instabilities or accuracy losses. We take the initial data $p_0=q_0=2$ and evolve the system for an overall time $T=100$.  

We consider the 3rd order stochastic approximation (\ref{Langevin1}), as well as its truncations to 2nd and 1st order obtained by removing the stochastic variables $\zeta_3$ (for 2nd order) or both $\zeta_3$ and $\zeta_2$ (for 1st order). We will denote the order of the stochastic approximation by $\gamma$. The ODEs in each time step are solved with the operator-splitting techniques using the PQ splitting (\ref{PQ}). We take the splitting order $s$ equal to the stochastic order, $s=\g$, or one unit higher, $s=\g+1$.  

Let us check that the 2nd and 3rd order schemes produce the solution which converges to the true solution of the SDE (\ref{Langevin0}). 
Since the true solution is not known analytically, we must find the proxy for it numerically.
First, we generate the white noise on a fine grid with step $h_*=2\cdot 10^{-5}$.
Then we solve \cref{Langevin0} numerically with this white noise using the scheme with $\g=1$, which is nothing but the well-known Euler--Maruyama method.
The obtained numerical solution is used as the proxy for the true solution.

Next, we adopt a grid with $h>h_*$ and compute the noise variables (\ref{newrandvars}) corresponding to the above realization of the white noise, as explained in Appendix~\ref{app:strong}.
We solve \cref{Langevin0} using the 2nd and 3rd order schemes and compare with the proxy solution by measuring the strong error at the final time $T$:
\be 
|\Delta z_T|\equiv\sqrt{|p(T)-p^{\rm true}(T)|^2+|q(T)-q^{\rm true}(T)|^2} \;.
\ee
In the end we average the measured value of the difference over 10 realizations of the noise. 

\begin{figure}[t!]
    \begin{minipage}[h]{0.49\linewidth}
        \centering    
        \includegraphics[width=1.0\linewidth]{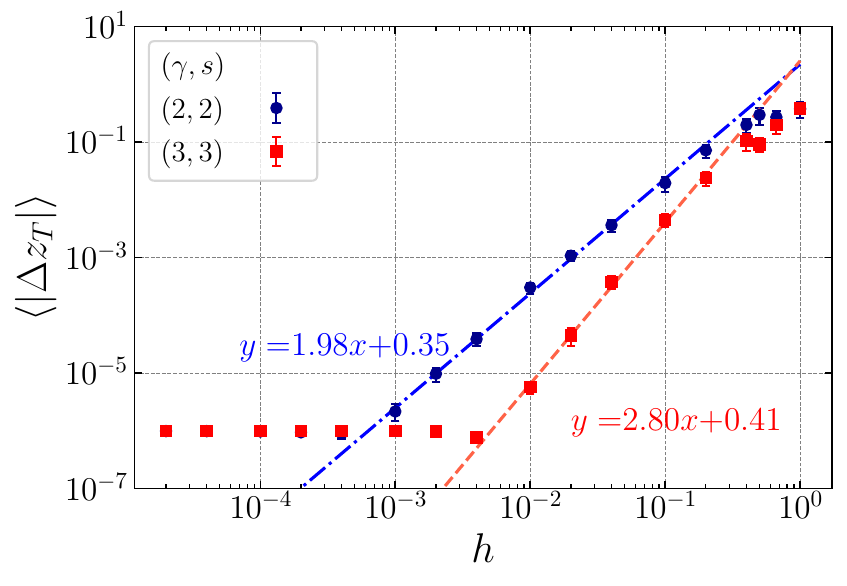}
	\end{minipage}~~
    \begin{minipage}[h]{0.49\linewidth}
        \centering     
        \includegraphics[width=1.0\linewidth]{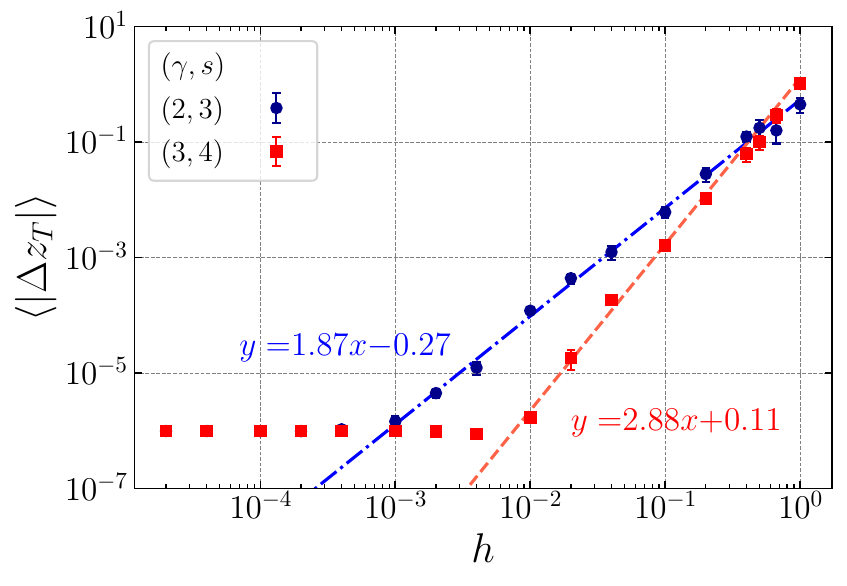}
	\end{minipage} 
    \caption{Difference between the numerical solutions obtained using higher-order schemes with time step $h$ and the reference solution computed with the 1st order scheme at a fine grid with $h_*=2\cdot 10^{-5}$. Blue circles (red squares) show the results for the schemes of stochastic order $\g=2$ ($\g=3$). The order of splitting is equal to $s=\g$ ($s=\g+1$) in the {\it left} ({\it right}) panel. The results are averaged over 10 realizations of the noise. The statistical uncertainty of the average is shown with the errorbars, which are barely visible in the plots. Straight lines represent fits to the data in the region where they follow a power-law dependence. }
\label{fig:pend_ref}
\end{figure}

The results of this procedure are shown in 
Fig.~\ref{fig:pend_ref}. 
The left panel shows the results for the schemes with the stochastic order $\g=2$ or $\g=3$ and equal splitting order, $s=\g$. Whereas in the right panel the splitting order is one unit higher, $s=\g+1$. We observe that all schemes exhibit convergence to the reference solution as $h$ decreases. The numerical schemes are stable at all values of $h$ we explored, up to $h=1$. Further, the scaling of the numerical error with $h$ is well fitted by power-law with the power matching the stochastic order,
\be 
\label{powerlaw}
\ev{|\Delta z_T|} \propto h^\g \;,
\ee 
down to the values $\ev{|\Delta z_T|} \sim 10^{-6}$. The convergence rate is independent of the splitting order $s$, which is expected since we take it to be larger or equal to the stochastic order. Still, taking $s=\g+1$ has a slight numerical advantage over the $s=\g$ case: For the same value of $h$, the error of the $s=\g+1$ schemes is about three times lower than the error of the $s=\g$ schemes.  

\begin{figure}[t!]
    \begin{minipage}[h]{0.49\linewidth}
        \centering
        \includegraphics[width=1.0\linewidth]{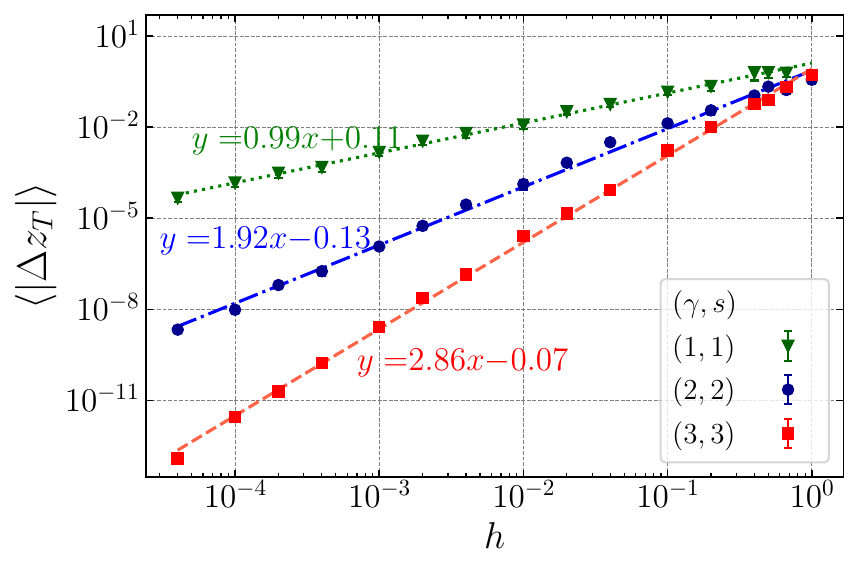}
	\end{minipage}~~
    \begin{minipage}[h]{0.49\linewidth}
        \centering
        \includegraphics[width=1.0\linewidth]{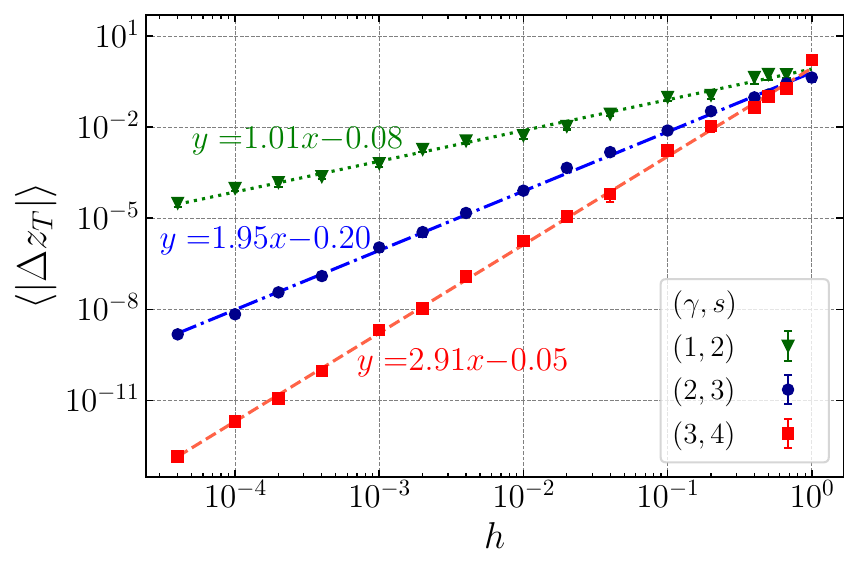}
	\end{minipage} 
    \caption{Scaling of the numerical error with the grid step $h$ for various schemes. The error is computed as the difference between the solution obtained using a given scheme and the reference solution obtained with the same scheme and the same realization of the white noise on a fine grid with $h_*=2\cdot 10^{-5}$. The measured error is averaged over 10 realizations of the white noise.
    Green triangles, blue circles and red squares show data points for the schemes with stochastic order $\gamma=1$, $2$, and $3$, respectively. {\it Left} ({\it right}) panel corresponds to the schemes with the splitting order $s=\g$ ($s=\g+1$). Straight lines show power law fits to the data.}
    \label{fig:pend_self}
\end{figure}

At small $h$ the data points in Fig.~\ref{fig:pend_ref} reach the floor at $\ev{|\Delta z_T|}\sim 10^{-6}$. This does not signal a loss of precision by the higher-order schemes. Rather, this is a consequence of the inaccuracy of our fiducial solution which itself was obtained numerically, using only 1st order scheme. To test the convergence rate at higher precision, we need a more accurate fiducial solution. This can be generated using the higher-order schemes themselves, since we have already established that they converge to the right answer. In Fig.~\ref{fig:pend_self} we plot the discrepancy between the solutions obtained with a given scheme for different values of $h$ and the reference solution obtained with the same scheme and the same realization of white noise on the fine grid with $h_*=2\cdot 10^{-5}$. For all schemes we see a perfect power-law scaling of the error consistent with the stochastic order, eq.~(\ref{powerlaw}). 
This confirms, in particular, the 3rd strong order convergence of the approximation (\ref{Langevin1}). 
We also see that the numerical prefactors in the errors of all schemes are order-one. Thus, increasing the strong order allows one to achieve much higher precision\footnote{We have probed $\ev{|\Delta z_T|}$ down to the values $\sim 10^{-12}$ at $h\sim 10^{-4}$. At this level of precision the numerical results get contaminated by the round-off errors if one uses the standard 64-bit arithmetic corresponding to the `double' type in C++. To avoid this problem, we used the 128-bit arithmetic implemented by the type `float\underline{ }128' in C++; see Appendix~\ref{app:strong} for details.}  
for the same time step $h$.
This gives computational advantage whenever the accuracy is required at the level of individual realizations of the stochastic force. 

\subsection{Stochastic anharmonic oscillator}
\label{ssec:osc}

Our next example is a one-dimensional oscillator with quartic anharmonicity. 
The potential and the corresponding force read,
\be \label{Pot_osc}
U(q)=\frac{q^2}{2}+\frac{q^4}{4} ~~~~\Longleftrightarrow~~~~
f(q)=-q-q^3\;.
\ee 
As in the previous subsection, we couple it to a heat bath with unit temperature, $\T=1$, dissipative coefficient $\eta=10$, and the noise amplitude $\s=\sqrt{2\eta\T}$. The initial conditions are $p_0=q_0=2$ and the system is evolved for a total time $T=100$. The model is a prototype of the classical lattice field theory studied below in Sec.~\ref{sec:field}. We use it to test the combination of the stochastic approximation (\ref{Langevin1}) with the LN splitting (\ref{LN}), where we take $f_{\rm lin}(q)=-q$ and 
$f_{\rm int}(q)=-q^3$. We also run simulations with the PQ splitting (\ref{PQ}) for comparison.  

\begin{figure}[t]
    \begin{minipage}[h]{0.49\linewidth}
        \centering    
        \includegraphics[width=1.0\linewidth]{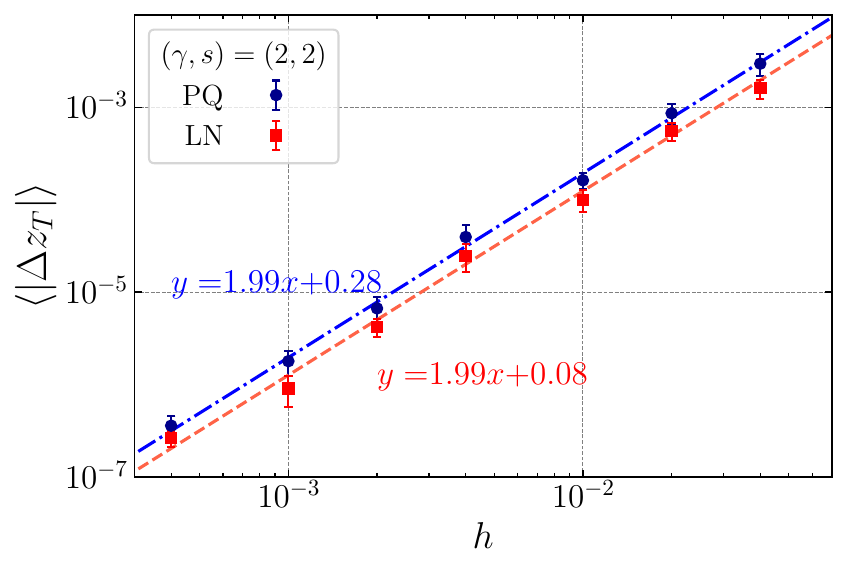}
	\end{minipage}~~
    \begin{minipage}[h]{0.49\linewidth}
        \centering       
        \includegraphics[width=1.0\linewidth]{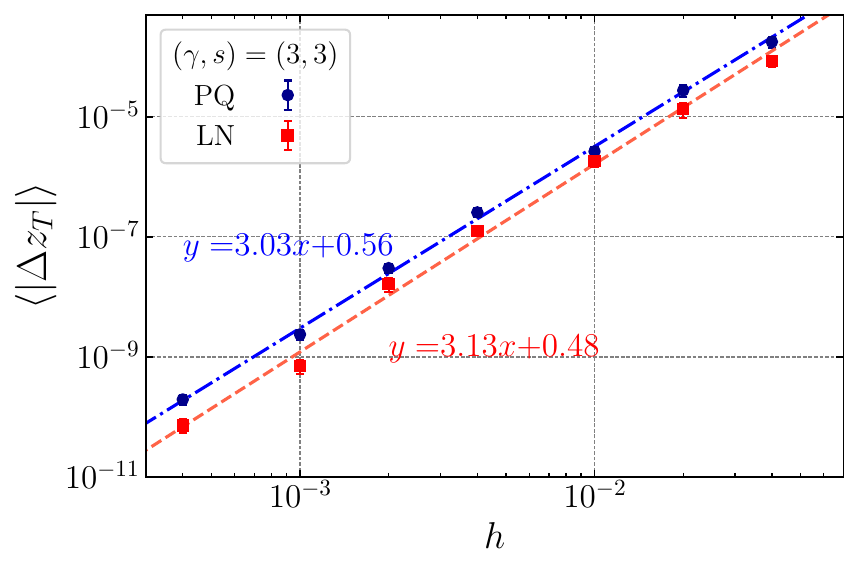}
	\end{minipage} 
    \caption{Scaling of the numerical error with the time step $h$ for various operator-splitting schemes applied to the quartic anharmonic oscillator. {\it Left panel:} 2nd order stochastic schemes with 2nd order splittings ($\g=s=2$). Blue circles (red squares) show the results obtained using PQ (LN) splitting. Straight lines show power-law fits to the data. The measurements are averaged over 10 realizations of the white noise. {\it Right panel:} same for the 3rd order schemes ($\g=s=3$). Note the different scales on the vertical axes in the two panels.
    }
	\label{fig:osc_self}
\end{figure}

As in the case of the stochastic pendulum, we made sure that both splittings lead to stable schemes converging to the correct continuum limit. We do not describe these tests in detail and focus on the measurements of the strong convergence order. We consider schemes with the splitting order equal to the stochastic order, $s=\g$. We use the same method as in Sec.~\ref{ssec:pend}, evaluating the difference between the numerical solutions with different time steps $h$ and the reference solution obtained using the same scheme with $h_*=2\cdot 10^{-5}$. The results are shown in Fig.~\ref{fig:osc_self} for the $\g=2$ schemes (left) and $\g=3$ schemes (right). The two sets of data points in each plot correspond to the PQ and LN splittings. We see that all schemes exhibit the expected order of strong convergence. Numerically, the LN schemes have slightly smaller errors (by a factor $\sim 2$), but the difference is rather mild. 

Confirmation of the strong convergence order of the stochastic schemes with LN splitting is important for the next section where we apply them to field theory. Note that a direct test of the strong convergence in the field theory setting is unfeasible due to the computer time and memory limitations.

\section{Field theory applications}
\label{sec:field}

We now apply the numerical scheme presented above to the stochastic field equation (\ref{LangEq}). 
We recall that this equation describes a scalar field in $(1+1)$-dimensional spacetime coupled to a heat bath. Te strength of coupling to the heat bath is characterized by the dissipation parameter $\eta$. The conservative part of the force is obtained from the potential density
\be 
\label{S}
V(\phi)=\frac{\phi^2}{2} + \frac{\a \phi^4}{4}\;.
\ee
We implement the pseudo-spectral method with LN splitting described in Sec.~\ref{ssec:fieldtheory}. 
An additional advantage of using the LN splitting is that in the examples below the nonlinear forces have a small (perturbative) effect on the evolution of the physical observables of interest.
Throughout this section we use the 4th order splitting, $s=4$, 
as a benchmark choice for the high-order ODE integrator, which is optimal for the accuracy and the number of operations.
We then switch the stochastic order $\g$ between 2 and 3, in order to see if the 3rd strong order scheme is more accurate when tested on the observables of physical interest.
We first study the theory with repulsive interaction, $\a=1$, which possesses stable ground state at $\phi=0$. Starting from the ground state, we let the system approach thermal equilibrium and investigate the precision with which the numerical code reproduces the key thermodynamic quantities. We then consider the theory with metastable vacuum, $\a=-1$, and obtain the false vacuum activation rate as function of $\eta$. The latter results were partially reported in \cite{Pirvu:2024ova,Pirvu:2024nbe}. In this work we extend them to higher values of $\eta$ and explore the high-$\eta$ asymptotics of the rate.

We work on a spatial lattice of length $L=100$ with periodic boundary conditions. The number of sites is 
$N=8192=2^{13}$, which corresponds to the lattice spacing $\ell = L/N = 1.22\cdot 10^{-2}$. We have checked that none of our results depend on the lattice parameters. We choose the temperature of the heat bath to be $\T=0.1$. This choice makes thermal effects significant but still allows us to treat them perturbatively when comparing the numerical results to the analytical predictions.  
Note that the thermodynamic properties of the system, such as $\T$, are 
determined by high-momentum Fourier modes of $\phi$, which are almost free and which are, therefore, evolved almost exactly by the pseudo-spectral LN splitting algorithm.

\subsection{Repulsive interaction: Weak convergence of thermodynamic observables}
\label{ssec:rep}

In the runs of this subsection the dissipation coefficient is fixed 
$\eta=1$. In this regime both the nonlinear 
Hamiltonian evolution of the system and stochastic effects are important.
For every choice of $\g=2,3$ and the time step $h$ we run a suite of $10^4$ simulations starting from the vacuum initial conditions $\phi=\pi=0$. The system is evolved for the total time $T=520$. It takes the time $t_{\rm th}\sim\eta^{-1}$ for the system to reach local thermal equilibrium.\footnote{As discussed in \cite{Pirvu:2024nbe}, the thermalization time of the system is set by $\eta^{-1}$ as long as $\eta\lesssim 1$. If $\eta$ is small, the thermalization time is determined by the self-interaction of the field and is very long, $t_{\rm th}\sim \T^{-4}$. Also, if $\eta$ is large, the field modes evolve slower because of overdamping, leading to $t_{\rm th}\sim \eta$. 
}
We let the system evolve for $20/\eta$ before taking measurements. We measure the power contained in the Fourier modes of the field and momentum, $|\tilde\phi_j|^2$ and $|\tilde\pi_j|^2$. We take 50 samples of the power spectra during each simulation, uniformly spanned between $20/\eta$ and $T$. The samples are separated by $10/\eta$ to avoid correlation between individual measurements. 
The total number of samples for fixed $\g$ and $h$ 
is therefore $5\cdot 10^5$. 

We compare the measurements with the theoretical expectations for 14 values of the time step in the interval $5\cdot 10^{-3}\leq h\leq 0.7$. In this way we infer the weak convergence order of these observables.
Recall that a numerical scheme is said to have weak convergence order $\beta$ with respect to an observable $X$ if the difference between the numerical average of this observable and its true expectation value scales as
\be
\label{weakX}
|\langle X\rangle_h-\langle X\rangle|\leq C_X\,h^\beta\;,
\ee
for an $h$-independent constant $C_X$ \cite{kloeden2013numerical}.

\subsubsection{Effective temperature}
\label{ssec:temp}

First, we measure the temperature of the thermalized field. By the equipartition theorem, the temperature is related to the momentum power spectrum:
\be\label{T1}
\ev{\left|\tilde\pi_j\right|^2} = \frac{\T}{\ell} \;.
\ee
We introduce the notion of an `effective temperature' of Fourier modes within a certain range of wavenumbers $k_{\rm min}\leq k_j\leq k_{\rm max}$. It is defined through the momentum power spectrum in the range,
\be \label{Teff}
\Teff = \ell \ev{\left|\tilde\pi_j\right|^2}_{k_{\rm min}\leq k_j\leq k_{\rm max}} \;,
\ee
where the averaging is performed both over the modes and over an ensemble of systems.  
The effective temperature quantifies how much the ensemble deviates from thermal equilibrium.
In our case the deviation is caused by the numerical errors introduced by the scheme,
and we will use $\Teff$ as a test function to measure the weak convergence. Specifically, we will
measure the relative difference $\delta_\T = |\Teff-\T|/\T$ as function of $h$.

We build an estimator of $\Teff$ for a given sample,
\be 
\label{Teffest1}
\mathcal{T}_{\rm eff}^{(J)} = \frac{\ell}{j_{\rm max}-j_{\rm min}+1}\sum_{j=j_{\rm min}}^{j_{\rm max}} \big|\tilde \pi_j^{(J)}\big|^2 \;,
\ee
where $J$ is the number of the sample and $\tilde\pi_j^{(J)}$ are the amplitudes of the momentum modes in this sample. Then we average over $M$ different samples and obtain an estimator of $\Teff$ for the ensemble,
\be 
\label{Teffest2}
\bar{\mathcal{T}}_{\rm eff}= \frac{1}{M}\sum_{J=1}^{M}\mathcal{T}_{\rm eff}^{(J)} \;.
\ee 
Due to the statistical nature of the ensemble and finite number of samples, the estimator has a non-zero variance computed as
\be \label{T_var}
\Delta_{\mathcal{T}} = \frac{1}{M-1}\sqrt{\sum_{J=1}^{M}\Big(\bar{\mathcal{T}}_{\rm eff}- \mathcal{T}_{\rm eff}^{(J)} \Big)^2} \;.
\ee 

\begin{figure}[t]
    \begin{minipage}[h]{0.49\linewidth}
        \centering
        \includegraphics[width=1.0\linewidth]{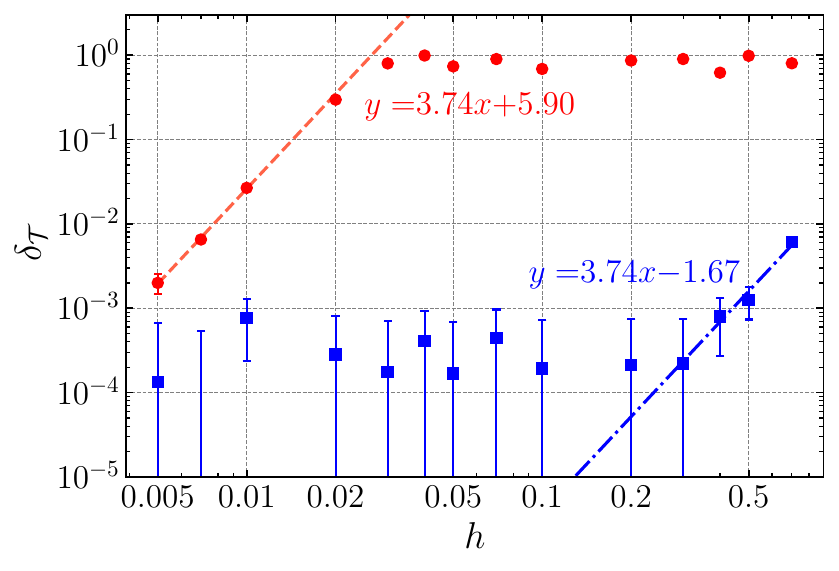}
	\end{minipage}~~
    \begin{minipage}[h]{0.49\linewidth}
        \centering
        \includegraphics[width=1.0\linewidth]{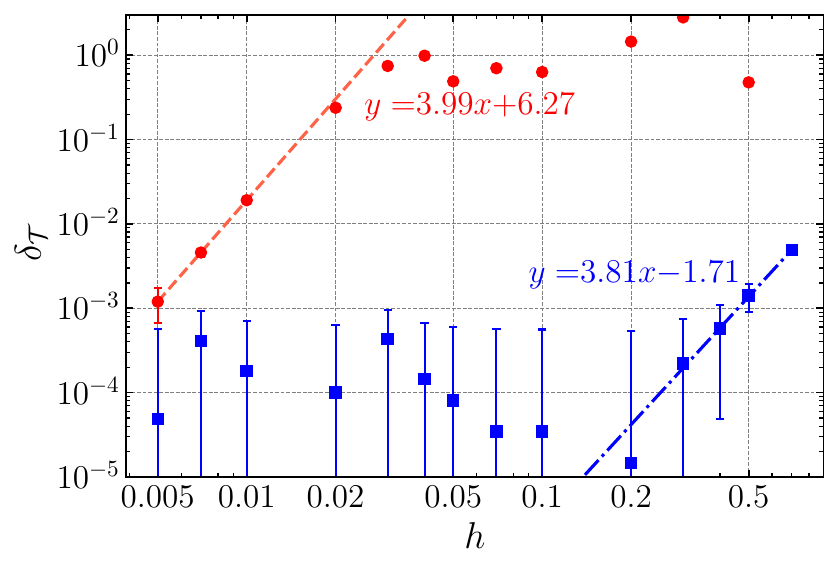}
	\end{minipage}
    \caption{Relative error in the effective temperature (\ref{Teff}) of short (red dots) and long (blue squares) Fourier modes of the field as function of the time step $h$. The effective temperature is measured from a suit of simulations evolved using the stochastic pseudo-spectral schemes with 4th order LN splitting. The stochastic order of the scheme is $\g=2$ ({\it left}) and ${\g=3}$ ({\it right}). Errorbars represent the statistical uncertainty of the measurements estimated using eq.~(\ref{T_var}). The red dashed and the blue dash-dotted lines are fits to the data points in the regions where they exhibit power-law scaling. 
    }
	\label{fig:temp23}
\end{figure}

The relative weak error $\delta_\T=|\bar{\T}_{\rm eff}-\T|/\T$ measured in simulations with different $h$ is shown in Fig.~\ref{fig:temp23} for the $\g=2$ scheme ({\it left}) and $\g=3$ ({\it right}). We use two sets of modes: long modes with order-one momenta, 
\be
\label{longk}
k_{\rm min}=0~,\qquad k_{\rm max}=30\cdot(2\pi/L)\simeq 1.88\qquad\qquad\text{(long modes)},
\ee
and short modes with momenta of order the inverse of the lattice spacing,
\be
\label{shortk}
k_{\rm min}=\pi/\ell-30\cdot(2\pi/L)\simeq 255.5~,\qquad k_{\rm max}=\pi/\ell\simeq 257.4
\qquad\qquad\text{(short modes)}.
\ee 
The relative error in the effective temperature of the long modes is shown by blue squares. We see that it is very small (below $10^{-3}$) for both schemes already at relatively large step $h\lesssim 0.4$. At $h<0.2$ it is completely buried below the statistical variance of the estimator. The relative error for short modes is shown by red dots. These require much smaller time steps $h\lesssim 0.02$ to be reproduced accurately. This is not surprising since these modes have very high frequencies, $\Omega_{\rm short}\sim 160$. However, once $h$ gets below the critical value, the relative error rapidly decreases and reaches down to $\delta_\T\sim 10^{-3}$ at $h\simeq 5\cdot 10^{-3}$. It is worth pointing out that despite the failure to reproduce the dynamics of short modes at $h> 0.02$, the numerical schemes do not exhibit any run-away behavior, i.e. they are stable for all time steps we explored. 

The power-law fits shown with the straight lines in Fig.~\ref{fig:temp23} suggest that the effective temperature converges with the 4th weak order for both $\g=2$ and $\g=3$ schemes. This may at first appear surprising. The reason for this behavior lies in the special properties of the momentum power spectrum. We see from eq.~(\ref{T1}) that it is insensitive to nonlinearities. In other words, the momentum power spectrum is the same as if we solved the linear equation obtained from (\ref{LangEq}) by setting $\a=0$. But for a linear equation, and for averages of observables quadratic in the fields, the order of the ODE approximation (\ref{Langevin2+}) gets enhanced. One can show using the expansions from Appendix~\ref{ssec:der} that the difference between the variances of the exact and approximate solutions in a single time step is 
\be
\big\langle\big(z(h)\big)^2\big\rangle-\big\langle \big(\hat z(h)\big)^2\big\rangle={\cal O}(h^5)\;.
\ee
This estimate is valid both for the $\g=3$ scheme and for its $\g=2$ truncation.\footnote{For the 1st order truncation the difference would be ${\cal O}(h^3)$ leading to a 2nd order weak convergence of the effective temperature.} 
Using then the same argument as in the proof of {\it Proposition 1} from Appendix~\ref{ssec:der}, one concludes that for a finite time interval the difference between the variances is ${\cal O}(h^4)$. 

To sum up, we have found that the $\g=2$ and $\g=3$ schemes lead to 4th order weak convergence of the effective temperature
because the latter does not feel the nonlinearities of the dynamics. We now consider a different observable which is sensitive to such nonlinearities.

\subsubsection{Field power spectrum}
\label{ssec:ps}

Our next example is the field power spectrum,
\be \label{ps}
\PP(k_j) = \ell\langle|\tilde\phi_j|^2\rangle \;.
\ee
It can be calculated perturbatively in the continuum limit using expansion in temperature $\T$; see Appendix~\ref{app:loops}. The expression including third-order corrections (3-loop effects) is,\footnote{While we focus on the repulsive interaction, $\a=1$, in this subsection, we write the expression for arbitrary values of $\a$ for generality.} 
\be \label{psTh}
\PP(k) = \T\left( k^2+1+\frac{3 \a\T}{2}-\frac{9\a^2\T^2}{8}-\frac{9\a^2\T^2}{2(k^2+9)}
+\frac{315}{128}\alpha^3 \T^3+\frac{27(5k^2+117)}{16(k^2+9)^2}\alpha^3 \T^3
\right)^{-1} \;.
\ee
Note that if we restrict to the correction linear in $\T$, we obtain $\PP(k)=\T (k^2+m_{\rm 1-loop}^2)^{-1}$, where 
$
m^2_{\rm 1-loop}=(1+3\a\T/2)
$
is the 1-loop thermal mass in the theory \cite{Pirvu:2024nbe,Boyanovsky:2003tc}. Beyond one loop, the corrections to $\PP(k)$ do not reduce to the thermal mass, but also include terms with modified momentum dependence.

For lattice measurements we take an aggregated power spectrum summed over the long modes (\ref{longk}),   
\be 
\label{Sdef}
{\cal S} = \frac{1}{j_{\rm max}+1}\,\sum_{j=0}^{j_{\rm max}} \PP(k_j)~,~~~~~j_{\rm max}=30 \;,
\ee
as a test function to probe the weak convergence.
Since the wavelengths of the modes are much longer than the lattice spacing, we can neglect here the corrections to the power spectrum due to the discreteness of the lattice and use eq.~(\ref{psTh}) for analytical predictions. 
By analogy with eqs.~(\ref{Teffest1}), (\ref{Teffest2}), we construct an estimator for this observable,
\be 
\label{Sest}
\bar {\cal S} = \frac{1}{M}\sum_{J=1}^M {\cal S}^{(J)} \;,\qquad~~~~
{\cal S}^{(J)} = \frac{\ell}{j_{\rm max}+1}\,\sum_{j=0}^{j_{\rm max}} \big| \tilde\phi_j ^{(J)} \big|^2 \;,
\ee
where $J$ labels the sampled field configurations and $M$ is the total number of samples. The variance of the estimator is computed as
\be \label{S_var}
\Delta_{\mathcal{S}} = \frac{1}{M-1}\sqrt{\sum_{J=1}^{M}\big(\bar{\mathcal{S}}- \mathcal{S}^{(J)} \big)^2} \;.
\ee 

\begin{figure}[t]
    \begin{minipage}[h]{0.49\linewidth}
        \centering
        \includegraphics[width=1.0\linewidth]{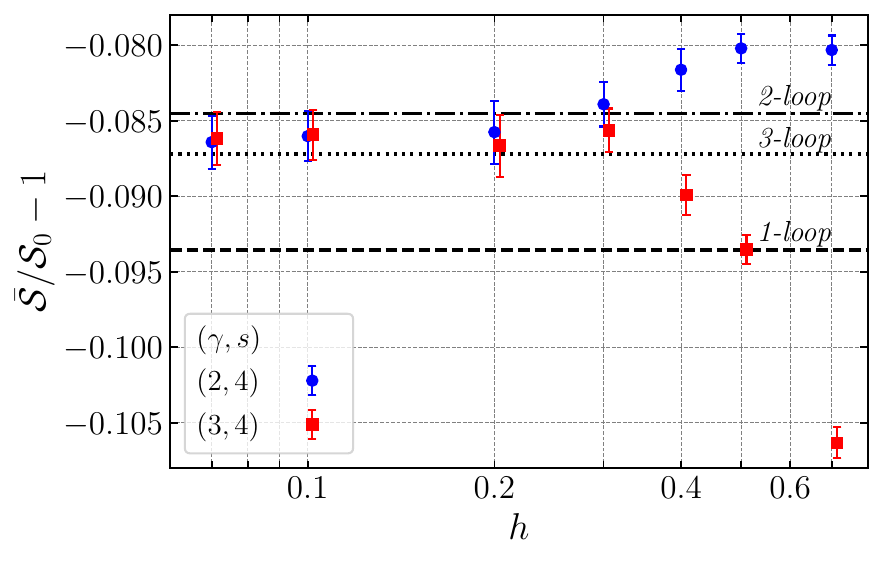}
	\end{minipage}~~
    \begin{minipage}[h]{0.47\linewidth}
        \centering
        \includegraphics[width=1.0\linewidth]{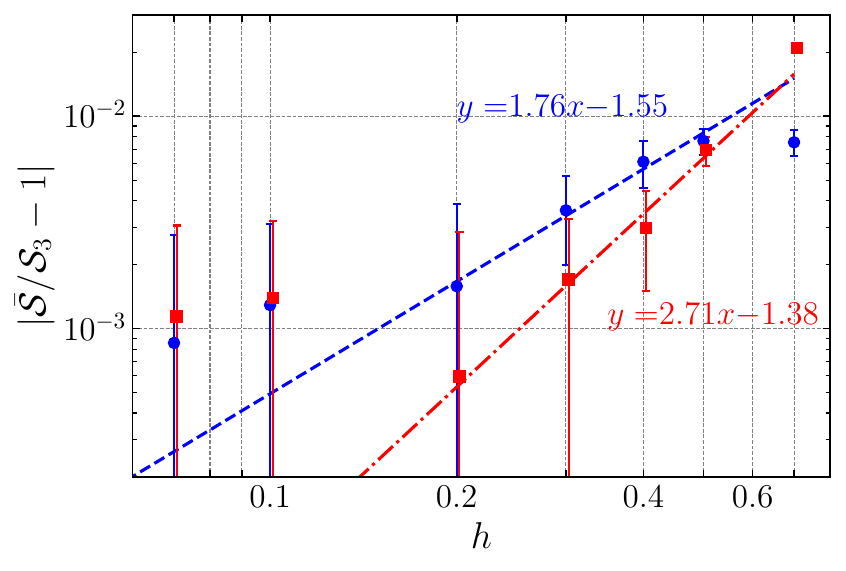}
	\end{minipage} 
    \caption{ \textit{Left:} Aggregated power spectrum of long field modes computed using the $\g=2$ (blue dots) and $\g=3$ (red squares) numerical schemes, as function of the time step $h$. Data points show the relative difference between the value measured 
    in the simulations and the tree-level analytical prediction. The horizontal dashed, dash-dotted and dotted lines correspond to the 1-, 2-, and 3-loop approximations, respectively. Errorbars represent the statistical uncertainty of the measurements estimated by eq.~(\ref{S_var}).  \textit{Right:} The error between the measured value and the 3-loop prediction. The straight lines show power-law fits to the data in the range $0.2\leqslant h\leqslant 0.5$. 
    }
	\label{fig:PS_new}
\end{figure}

In the left panel of Fig.~\ref{fig:PS_new} we present the value of $\bar{\cal S}$ measured in the simulations using the $\g=2$ and $\g=3$ schemes with different time steps. The measured value is normalized to the tree-level prediction ${\cal S}_0= (j_{\rm max}+1)^{-1}\sum_j \T(k_j^2+1)^{-1}$. The horizontal lines in the plot show the predictions including 1-, 2-, and 3-loop corrections. We see that when $h$ decreases, the numerical results converge to the value consistent with the 3-loop result. They are clearly incompatible with the tree-level and 1-loop approximations, and mildly disfavor the 2-loop value. 
The convergence is fast: the relative weak error of the simulated result with respect to the 3-loop value is smaller than the statistical uncertainty, $|\bar{\cal S}/{\cal S}_3-1|\lesssim 10^{-3}$, 
already at $h\lesssim 0.2$. This is clearly seen in the right panel of Fig.~\ref{fig:PS_new} showing this error in the log-scale. Note that this time step is still much larger than the period of the shortest modes on the lattice and, as we saw in the previous subsection, is not sufficient to correctly reproduce their temperature. Nevertheless, this inaccuracy in simulating the short modes does not affect the long-modes' power. From the right panel in Fig.~\ref{fig:PS_new} we also see that the two numerical schemes used in the simulations provide weak convergence of the observable ${\cal S}$ at the rate close to their 
nominal orders $2$ and $3$. This is reassuring, confirming that both schemes work as expected.     

The results of this subsection provide evidence that the $\g=2$ and $\g=3$ schemes correctly reproduce the statistical properties of the field sensitive to nonlinear dynamics. It is worth emphasizing that, in order to match the accuracy of the numerics, we had to go up to the 3-loop corrections in the analytical expression for the power spectrum. Furthermore, we note that the numerical result in the left panel of Fig.~\ref{fig:PS_new} slightly deviated upward from the 3-loop value at small $h$ (though still within the statistical uncertainty). This agrees with the expected sign of the 4-loop corrections omitted in this work.

\subsection{Attractive interaction: Activation of the metastable vacuum}
\label{ssec:decay}

Having tested the numerical schemes on the examples where the analytical answer is known with high precision, let us turn to a situation in which numerical simulations bring new insight. We consider eq.~(\ref{LangEq}) with negative parameter $\a=-1$. The corresponding potential (\ref{S}) is not bounded from below and hence the vacuum $\phi=0$ is metastable. Thermal fluctuations classically activate its decay, resulting in a run-away behavior $\phi\to\pm\infty$. The observable we study in this subsection is the decay rate $\Gamma$. 

The standard approach to the false vacuum decay at finite temperature treats $\Gamma$ as a thermodynamic quantity and relates it to the imaginary part of the false vacuum free energy \cite{Linde:1980tt,Linde:1981zj,Affleck:1980ac,Weinberg:2012pjx}. If the coupling to the external heat bath can be neglected (i.e. $\eta=0$), one can use the thermal field theory techniques and extract the free energy from the Euclidean path integral representation for the partition function. Application of this method to the field theory with potential (\ref{S}) gives \cite{Pirvu:2024ova,Pirvu:2024nbe},
\be
\label{G_E2}
\Gamma_{\rm E}(\T)=\frac{6}{\pi}\, \sqrt{\frac{E_b}{2\pi \T}}\;\e^{-E_b/\T}   \;,
\ee 
where $E_b=4/3$ is the energy of the critical bubble --- an unstable static field configuration separating the metastable vacuum from the run-away region. 

The dissipative and stochastic effects introduced by the coupling to the bath are taken into account using methods of classical statistical mechanics \cite{Langer:1969bc} and lead to an extra factor in the decay rate depending on the dissipative parameter in the Langevin equation (\ref{LangEq}), 
\be 
\label{pref_Langer}
\Gamma_{\rm stat}(\eta, \T) = \left(\sqrt{1+\frac{\eta^2}{4\omega_-^2}} -
\frac{\eta}{2\omega_-}\right) \Gamma_{\rm E}(\T) \;,
\ee
where $\omega_-$ is the critical bubble growth rate (i.e. the growth rate of its unstable mode); in the theory at hand $\omega_-=\sqrt{3}$ \cite{Pirvu:2024ova,Pirvu:2024nbe}.

The expressions (\ref{G_E2}), (\ref{pref_Langer}) are leading-order results and are expected to receive corrections. While the Euclidean formalism in principle allows one to systematically obtain the so-called `statistical' corrections coming from the expansion of free energy in powers of $\T/E_b$, there is no established procedure to evaluate corrections to the rate coming from the dynamics of bubble nucleation; see e.g. discussion in \cite{Gould:2021ccf}. In particular, it is unclear if the assumption of thermal equilibrium built in the derivation of eq.~(\ref{G_E2}) is justified during the bubble nucleation process. Further, the derivation of eq.~(\ref{pref_Langer}) is valid only for strong enough coupling to the bath \cite{Hanggi:1990zz,Pirvu:2024nbe},
\be
\label{smalleta}
\eta > \frac{\omega_-\T}{E_b}\;.
\ee
We have recently reported \cite{Pirvu:2024ova,Pirvu:2024nbe} large deviations from eqs.~(\ref{G_E2}), (\ref{pref_Langer}) observed in real-time numerical simulations of false vacuum activation using the Langevin equation (\ref{LangEq}). In these works we focused on the regime of zero or weak dissipation violating the condition (\ref{smalleta}). We also observed that the rate is closer to the prediction (\ref{pref_Langer}) when $\eta$ gets of order one or larger. Here we study the regime $\eta\gg 1$ in more detail, using it as a playground for application of the numerical methods developed in the current work. 

\begin{figure}[t]
\begin{minipage}[h]{0.49\linewidth}
        \centering
        \includegraphics[width=1.0\linewidth]{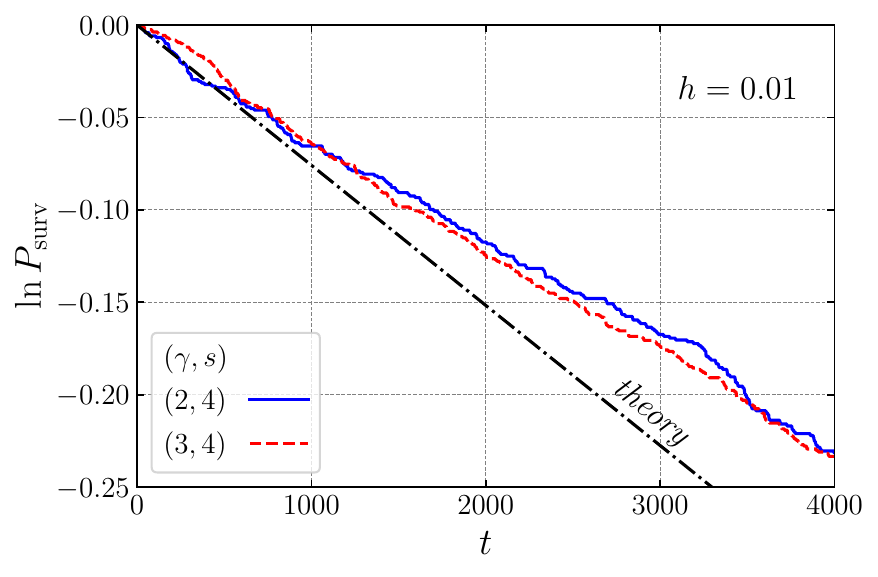}
	\end{minipage} ~~
    \begin{minipage}[h]{0.47\linewidth}
        \centering
        \includegraphics[width=1.0\linewidth]{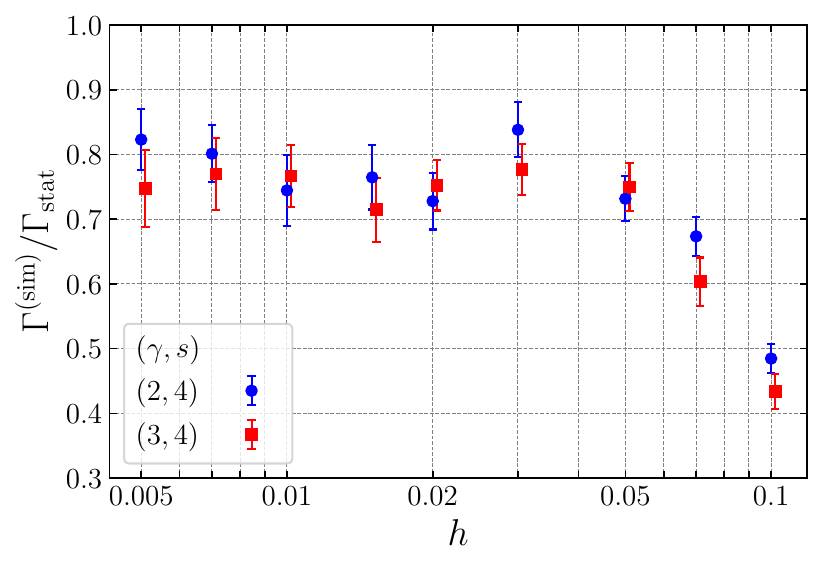}
	\end{minipage}
    \caption{{\it Left:} Survival probability of the false vacuum as function of time in simulations using the $\g=2$ (blue) and $\g=3$ (red) stochastic schemes with time step $h=0.01$. The parameters of the simulations are $\T=0.1$, $\eta=10$. Black dash-dotted line shows the leading-order theoretical prediction (\ref{pref_Langer}).  
    {\it Right:} The ratio of measured decay rate to the leading-order theoretical prediction as function of the time step. Other parameters are the same as in the left panel. Errorbars show the statistical uncertainty of the measurement.
    }
	\label{fig:pref}
\end{figure}

We measure the decay rate as follows. We prepare a suit of simulations with initial conditions close to the metastable vacuum and evolve them using eq.~(\ref{LangEq}). If a given simulation enters into run-away, we identify it as decayed and eliminate from the ensemble. At regular time intervals we count the number of remaining simulations. Its ratio to the initial number of simulations gives the survival probability $P_{\rm surv}(t)$. A typical behavior of this observable is shown in the left panel of Fig.~\ref{fig:pref}. Barring possible transient effects in the beginning of simulations, we fit the probability with the exponential law,   
\be \label{Psurv}
\ln P_{\rm surv}(t)={\rm const}-\Gamma\cdot Lt \;,
\ee
where $L$ is the spatial extent of the lattice, and $\Gamma$ is the sought-after decay rate. The relative statistical uncertainty in measurements of $\Gamma$ follows the Poisson statistics, $\Delta_\Gamma/\Gamma=N_{\rm dec}^{-1/2}$, where $N_{\rm dec}$ is the number of decays registered during the simulation time. The reader is referred to \cite{Pirvu:2024nbe} for more details on the measurement procedure. 

In the first set of simulations we fix the physical parameters $\T=0.1$, $\eta=10$ and study the numerical convergence of the rate by varying the time step $h$. We explore the schemes with the stochastic orders $\g=2$ and $\g=3$. In these runs we start from vanishing initial conditions $\phi=\pi=0$ and allow the field to thermalize due to the interaction with the bath. Since thermalization rate is much faster than $\Gamma$, the choice of initial conditions does not affect the measurement. 

\begin{figure}[t]
\centering
\begin{tikzpicture}

\node[anchor=south west,inner sep=0] (image) at (0,0) {\includegraphics[width=0.48\linewidth]{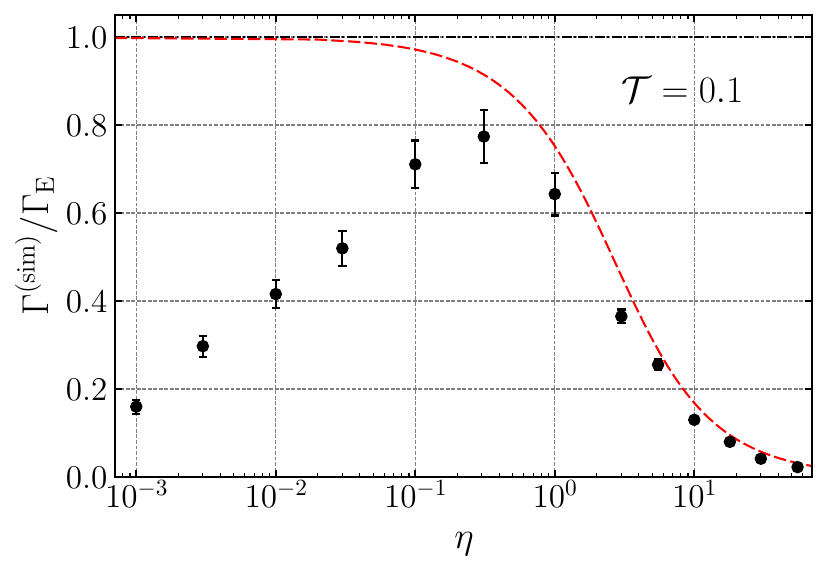}};
\node[anchor=south west,inner sep=0] (image) at (8.4,0) {\includegraphics[width=0.49\linewidth]{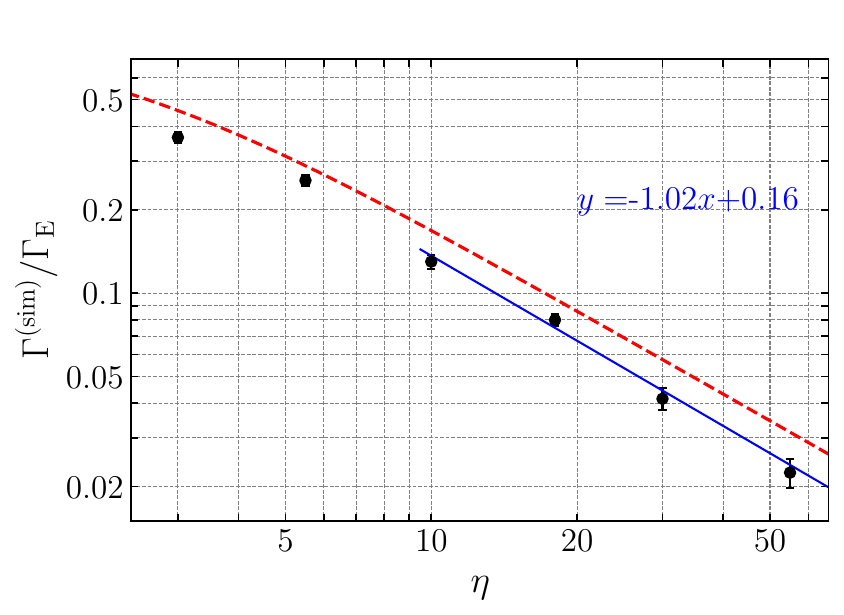}};
\draw[gray,very thick] (5.7,0.95) rectangle (7.8,3.8);
\draw[gray,very thick] (7.8,0.95) -- (9.7,0.92);
\draw[gray,very thick] (5.7,3.8) -- (9.7,5.3);

\end{tikzpicture}
\caption{{\it Left:} Ratio of the false vacuum decay rate measured in the simulations to the Euclidean expression (\ref{G_E2}) as function of the dissipation coefficient $\eta$ at fixed temperature $\T=0.1$. Errorbars show the statistical uncertainty of the measurements. Red dashed line shows the theoretical prediction (\ref{pref_Langer}). Simulations are performed using 3rd order stochastic pseudo-spectral scheme with 4th order operator-splitting; time step is $h=0.01$.   
{\it Right:} Zoom in on the region of strong dissipation. Straight blue line shows the power-law fit to the last four data points.}
\label{fig:decay_rate}
\end{figure}
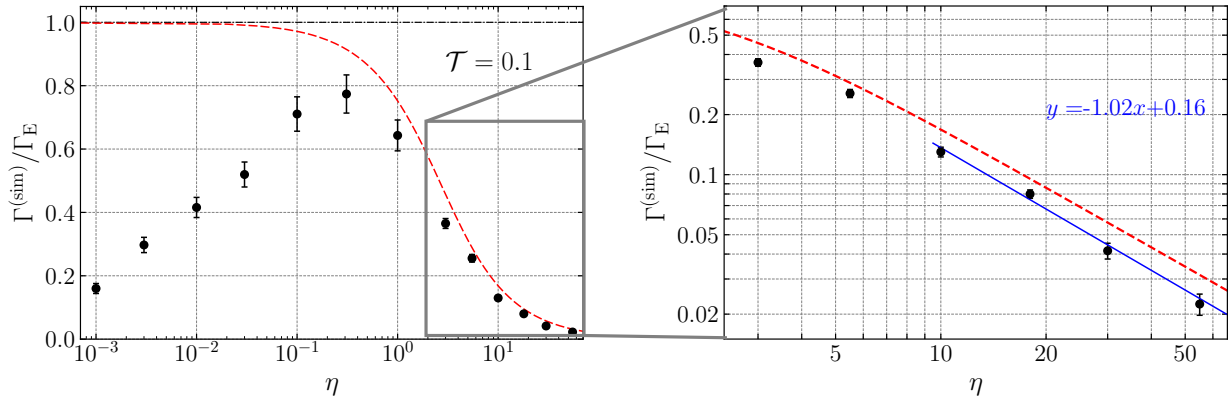

The ratio of the measured decay rate $\Gamma^{\rm (sim)}$ to the leading-order theoretical prediction (\ref{pref_Langer}) is shown in the right panel of Fig.~\ref{fig:pref}. Both schemes demonstrate convergence at $h\lesssim 0.05$ to the same value within the statistical uncertainty. In Sec.~\ref{ssec:rep} we saw that these time steps are sufficient to accurately reproduce the spectrum of long field modes, but may yet be too large to correctly capture the dynamics of the shortest modes. The latter, however, appear to be irrelevant for the false vacuum decay since the bubble nucleation is driven by modes with wavelengths of order unity \cite{Pirvu:2024nbe}. The statistical uncertainty of the measurement prevents us from seeing the difference between the 2nd and 3rd order schemes. Nevertheless, we expect (though we have not checked explicitly) that the $\g=3$ algorithm ensures higher precision at the level of individual field configurations, so we choose to use it in the rest of this section and adopt $h=0.01$ as the fiducial time step.

Figure~\ref{fig:pref} shows that the measured rate $\Gamma^{\rm (sim)}$ is systematically lower than the predicted value (\ref{pref_Langer}). The difference is significant. To explore it further, we perform simulations at other values of the dissipation coefficient $\eta$ keeping the temperature $\T=0.1$ fixed.\footnote{To suppress transients, we set the initial conditions in these runs according to the thermal distribution of fluctuations around the false vacuum; see \cite{Pirvu:2024nbe} for details.}
The measured decay rate normalized to the $\eta$-independent Euclidean prediction (\ref{G_E2}) is shown in Fig.~\ref{fig:decay_rate}. We see that the discrepancy becomes bigger at $\eta<0.03$, so that at $\eta=10^{-3}$ the measured rate is only $\sim 0.15\, \Gamma_{\rm stat}$. This suppression was first reported in \cite{Pirvu:2024ova,Pirvu:2024nbe} and was interpreted as a sign of violation of thermal equilibrium during the bubble nucleation. 

On the other hand, at $\eta\gtrsim 1$, the ratio between $\Gamma^{\rm (sim)}$ and $\Gamma_{\rm stat}$ stays approximately constant. 
This is clearly seen in the right panel of Fig.~\ref{fig:decay_rate} showing these two quantities (normalized to $\Gamma_{\rm E}$) in the logarithmic scale. In particular, we find that the measured rate is inversely proportional to the dissipation parameter, $\Gamma^{\rm (sim)}\propto \eta^{-1}$ at $\eta\gg 1$. This agrees well with the dependence predicted by eq.~(\ref{pref_Langer}). The coefficient in this dependence is, however, $\sim 20\%$ smaller than the theoretical prediction. This discrepancy is of the same order as expected ${\cal O}(\T)$ corrections to eq.~(\ref{pref_Langer}). We leave analytic derivation of these corrections to future work.

Before closing this section, let us mention the corrections to the measured decay rate due to the finite lattice size and spacing. These were analyzed in \cite{Pirvu:2024nbe} and are negligible for the lattice parameters adopted in our study.

\section{Discussion and conclusions}
\label{sec:disc}

We have presented a framework for constructing numerical schemes of up to 3rd strong convergence order for a class of stochastic differential equations (SDEs) with additive noise. A notable representative of this class is the (underdamped) Langevin equation which we used to illustrate the applications of our method. The method is equally suited for ordinary and partial differential equations of this type. In particular, it provides efficient algorithms to simulate real-time dynamics of classical fields on a lattice subject to thermal noise.

The key idea of the method is the separation of the numerical algorithm into two stages. In the first stage the SDE (\ref{NLstoch+}) with the singular white noise is substituted by a regular ODE (\ref{Langevin2+}) whose solution approximates the stochastic trajectory within a single time step. The stochasticity is encoded in the ODE coefficients which are expressed through a few independent Gaussian random variables sampled at each time step. In the second stage the resulting ODE is solved with a conventional ODE solver using the operator-splitting techniques. The choice of the ODE solver is independent of the first stage of the algorithm, the only requirement being that its order should not be lower than the order of the stochastic approximation. While the idea of approximating SDE with an ODE is not new, we believe our framework for the first time fully exploits its flexibility and makes the separation between the construction of the stochastic approximation and the choice of the operator-splitting method transparent. 

We have constructed several algorithms by combining stochastic approximations of different orders with various operator-splitting schemes and numerically tested 
their strong convergence. For this purpose, we considered two simple mechanical systems --- stochastic pendulum and stochastic anharmonic oscillator --- and observed that in both cases the stochastic schemes of 2nd and 3rd order yield dramatic gain of accuracy compared to the 1st order Euler--Maruyama method. 

When applied to field theory on a lattice, our approach naturally leads to pseudo-spectral algorithms using the splitting of the evolution operator into the linear and nonlinear parts. Notably, the part of the equation encoding the random force can be chosen to be constant over the time step and can be included into the linear evolution. The random variables entering into it are sampled in the Fourier space once in the beginning of the time step and remain fixed during the iterations of the splitting routine. 

We have implemented two algorithms of this type, of 2nd and 3rd strong order, in real-time simulations of a relativistic scalar field with quartic self-interaction in (1+1) dimensions coupled to an external heat bath. We let the field evolve from vanishing initial conditions and measured the effective temperature of its Fourier modes and their power spectrum. Both algorithms were found to reproduce these observables with high accuracy already at moderate values of the time step $h$. Namely, the relative numerical error dropped below the statistical uncertainty $\sim 10^{-3}$ due to the finite number of field samples used in the measurement at $\Omega h\lesssim 0.2$, where $\Omega$ is the frequency of the relevant modes. The algorithms were also found to be free from numerical instabilities for all values of the time step we explored, even when the product $\Omega h$ significantly exceeded unity for the shortest lattice modes.  

To illustrate the exploratory potential of the method, we applied it to the field theory with negative quartic coupling which possesses metastable vacuum state and measured the rate of its thermal activation. 
This required simulating the stochastic dynamics of the system on a large time interval and demonstrated the long-time stability of our high-order scheme.
The study complemented previous works \cite{Pirvu:2024ova,Pirvu:2024nbe} by accessing for the first time the regime of strong dissipation. We found the rate in this regime to be inversely proportional to the dissipation coefficient, in agreement with the theoretical expectations. However, the overall magnitude of the rate was found to be $\sim20\%$ lower than the value predicted by the leading-order theory. This indicates presence of higher-order corrections and motivates their analytical understanding.   
 
Is it possible to further improve the numerical scheme and reach strong convergence order higher than 3? The main obstruction in this way are non-Gaussian terms in the difference between the exact stochastic trajectory and the solution of the approximating ODE appearing at order ${\cal O}(h^4)$; see eq.~(\ref{diff4}). This prevents construction of stochastic schemes of 4th order without introducing non-Gaussian random variables. One may still increase the stochastic order of the scheme to $3.5$ with just one more Gaussian variable at each time step to cancel the first Gaussian term in the difference (\ref{diff4}). In addition, the approximating ODE needs to be modified to absorb the average of the non-Gaussian piece. It remains to be seen if this limited gain in convergence is enough to justify the complication of the numerical algorithm.  

We have proved the strong convergence of our schemes under certain technical assumptions on the deterministic force function in the Lagnevin equation. It would be interesting to see if the proof can be generalized to a broader class of force functions using, e.g., the approach of  \cite{CuiHongSheng}.

Our scheme can be extended beyond the Langevin equation (\ref{KG}) (e.g., by introducing time-dependent forces or nonlinear damping term) and to other stochastic partial differential equations admitting lattice discretization (e.g., by considering more general spatial derivative operators than in (\ref{LangEq})).
As long as these generalizations satisfy condition (\ref{Hess}), the 3rd strong order will be maintained.

It would be interesting to study the generalization of our method to SDEs with more general noise. 
Adaptation to the general L\'{e}vy-type noise \cite{platen2010numerical} appears straightforward: the approximating ODE would still have the form (\ref{Langevin2}), with the random parameters (\ref{newrandvars}). The statistics of the latter, however, will not, in general, be Gaussian, and must be computed from the statistical properties of the noise. Thus, the algorithm will have to be supplemented with an efficient way to sample these non-Gaussian variables.

Generalization to multiplicative noise appears more challenging. Beyond the 1st strong order, the approximating ODE in this case would involve random variables nonlinearly related to the noise functions and possessing complicated statistical properties.
It would be interesting to identify the conditions analogous to our \cref{Hess}, under which these nonlinear random variables could be simplified. We leave this study for the future.

\section*{Acknowledgments} 

We thank Mustafa Amin, Daniel G. Figueroa, David Del Rey Fernandez, Sita Gakkhar, Matthew Johnson, Rob Myers, Dalila 
P\^irvu and Jury Radkowski for useful discussions.
Research at Perimeter Institute is supported in part by the Government
of Canada through the Department of Innovation, Science and Economic
Development Canada and by the Province of Ontario through the Ministry
of Colleges and Universities. The work of SS is supported by the Natural Sciences and Engineering Research Council (NSERC) of Canada. This research was enabled in part by support provided by Compute Ontario (www.computeontario.ca) and Digital Research Alliance of Canada (alliancecan.ca).

\appendix

\section{Derivation of the approximating ODE}
\label{ssec:der}

In this Appendix we derive the approximating ODE (\ref{Langevin2+}) to the SDE (\ref{NLstoch+}) and analyze its convergence properties. 
We start by considering eq.~(\ref{NLstoch+}) in a small interval 
$0\leq t\leq h$ and rewrite it in the integral form,
\be\label{NLint}
z^i(t)=z^i_0+\int_0^t \diff t' \big[ F^i(z(t'))+\s^i_{\;r}\,\xi^r(t')\big]\;.
\ee
This equation can be solved iteratively. One starts with the zeroth-order approximation $z_0(t)=z_0$ and substitutes it to the r.h.s. In this way one obtains a first-order approximation $z_1(t)$. Substituting this again into the r.h.s.~and Taylor expanding the force term, one obtains $z_2(t)$, and so on. In performing the Taylor expansion one treats the noise $\xi(t)$ as a quantity of order $h^{-1/2}$, since its squared integral is (on average) of order one,
\be
\int_0^h \diff t'\,\langle \xi^r(t)\,\xi^s(t')\rangle=\delta^{rs}\;.
\ee
After three iterations we obtain the representation for the solution of (\ref{NLint}) through order $h^3$,
\be\label{NLtrue}
\begin{split}
z^i(t) &= z_0^i+ w_1^i(t)+ t \bar F^i+\bar F^i_{\;,j}w_2^j(t)
+\frac{t^2}{2}\bar F^i_{\;,j}\bar F^j\\
&+\bar F^i_{\;,j} \bar F^j_{\;,k}w_3^k(t)
+\bar F^i_{\;,jk} \bar F^j w_4^k(t)
+\frac{t^3}{6}(\bar F^i_{\;,j} \bar F^j_{\;,k}\bar{F}^k
+\bar F^i_{\;,jk} \bar F^j \bar F^k)
+z^i_{\rm NG}(t)+\mathcal{O}(h^{7/2})\;,
\end{split}
\ee
where we have denoted with bar the quantities evaluated at $z_0$, i.e. $\bar{F}^i = F^i(z_0), \bar F^i_{\;,j}=\left.\frac{\d F^i}{\d z^j}\right\vert_{z=z_0}, \bar F^i_{\;,jk}=\left.\frac{\d^2 F^i}{\d z^j\d z^k}\right\vert_{z=z_0}$ etc. We also introduced the stochastic integrals, 
\bseq
\label{stochints}
\begin{align}
&w_1^i(t)=\s^i_{\;r}\int_0^t\diff t' \xi^r(t')\;,&&
w_2^i(t)=\s^i_{\;r}\int_0^t\diff t' (t-t')\xi^r(t')\;,\\
&w_3^i(t)=\s^i_{\;r} \int_0^t \diff t' \frac{(t-t')^2}{2}\xi^r(t')\;,&&
w_4^i(t)=\s^i_{\;r} \int_0^t \diff t' \frac{t^2-t'^2}{2}\xi^r(t')\;.
\end{align}
\eseq
The term $z_{\rm NG}(t)$ stands for the non-Gaussian contributions,
\be
\label{trueNG}
\begin{split}
z_{\rm NG}^i(t)=&\frac{1}{2}\bar F^i_{\;,jk}\int_0^t\diff t'w_1^j(t')w_1^k(t')
+\frac{1}{6}\bar F^i_{\;,jkl} \int_0^t\diff t'w_1^j(t')w_1^k(t')w_1^l(t')\\
&+\frac{1}{2}\bar F^i_{\;,j} \bar F^j_{\;,kl} \int_0^t\diff t'(t-t')w_1^k(t')w_1^l(t')
+\frac{1}{2} \bar F^i_{\;,jkl} \bar F^j \int_0^t\diff t'\;t' w_1^k(t')w_1^l(t')\\
&+\bar F^i_{\;,jk}\bar F^k_{\;,l}\int_0^t\diff t'w_1^j(t')w_2^l(t')
+\frac{1}{24} \bar F^i_{\;,jklm}\int_0^t\diff t'w_1^j(t')w_1^k(t')w_1^l(t')w_1^m(t')\;.
\end{split}
\ee
Since $w_1$ is order $h^{1/2}$, the non-Gaussianities start at order $\mathcal{O}(h^2)$. Remarkably, they all vanish once we impose the condition (\ref{Hess}). In this case eq.~(\ref{NLtrue}) simplifies: the term with $w_4(t)$ and $z_{\rm NG}(t)$ drop out.
In case of the weaker condition (\ref{Hess1}), the non-Gaussianity appears first at order $\mathcal{O}(h^3)$ and corresponds to the first term in the last line of eq.~(\ref{trueNG}).

We want to compare (\ref{NLtrue}) to the solution of a regular ODE, 
\be\label{NLreg}
\dot{\hat z}^i= F^i(\hat z)+g_1^i+g_2^i(\hat z)+g_3^i(\hat z)\;
\ee 
where $g_1^i=h^{-1}w_1^i(h)$ is the time-averaged stochastic force, and  
$g_{2}$, $g_3$ are corrections of order $h^{1/2}$ and $h^{3/2}$, respectively. We assume that these corrections do not explicitly depend on time but allow them to smoothly depend on the coordinates. Writing an integral representation for this equation similar to (\ref{NLint}) and iterating it three times we obtain,
\be\label{NLreg3}
\begin{split}
\hat z^i(t)=&z_0^i+tg^i_1+t\bar F^i + \frac{t^2}{2}\bar F^i_{\;,j}g_1^j
+t \bar g_2^i
+\frac{t^2}{2}\bar F^i_{\;,j}\bar F^j
+\frac{t^2}{2}\bar g_{2,j}^i g^j_1\\
&+\frac{t^3}{6}\bar F^i_{\;,j}\bar F^j_{\;,k}g_1^k
+\frac{t^2}{2} \bar F^i_{\;,j}\bar g_2^j
+\frac{t^2}{2}  \bar g^i_{2,j}\bar F^j
+\frac{t^3}{6} \bar g^i_{2,jk}g_1^j g_1^k
+t \bar g_{3}^i\\
&+\frac{t^3}{6} (\bar F^i_{\;,j}\bar F^j_{\;,k}\bar F^k
+\bar F^i_{\;,jk} \bar F^j\bar F^k) 
+\frac{t^3}{6}\bar F^i_{\;,j}\bar g^j_{2,k} g_1^k
+\frac{t^3}{6}\bar g^i_{2,j}\bar F^j_{\;,k} g_1^k\\
&+\frac{t^2}{2}\bar g^i_{2,j}\bar g^j_{2}
+\frac{t^3}{3} \bar g^i_{2,jk}\bar F^j g_1^k
+\frac{t^2}{2} \bar g^i_{3,j} g_1^j
+\mathcal{O}(h^{7/2})\;.
\end{split}
\ee
Here we have used the property (\ref{Hess}) to simplify the expression.
We now adjust the functions $g_2$, $g_3$ in such a way 
that this expression matches the 3rd order approximation (\ref{NLtrue}) at $t=h$.
From matching the first lines in (\ref{NLtrue}) and (\ref{NLreg3}) we determine
\be\label{NLmatch2}
g_2^i(\hat z)=F^i_{\;,j}(\hat z)\left(\frac{1}{h}w_2^j(h)-\frac{1}{2}w_1^j(h)\right)\;.
\ee
Note that the condition (\ref{Hess}) implies that $g_2^i(\hat z)$ is actually constant, $g^i_{2,j}=0$. Then equating the rest of the expressions (\ref{NLtrue}), (\ref{NLreg3}) yields 
\be
\label{NLmatch3}
g_3^i(\hat z)=F^i_{\;,j}(\hat z)F^j_{\;,k}(\hat z)
\left(\frac{1}{h}w_3^k(h)-\frac{1}{2}w_2^k(h)+\frac{h}{12}w_1^k(h)\right).
\ee
Substituting here the stochastic integrals (\ref{stochints}), we see that eq.~(\ref{NLreg}) takes the form (\ref{Langevin2}) with the stochastic variables (\ref{newrandvars}). 

To study the accuracy of the approximation, we perform one more iteration of the solutions. 
After a tedious but straightforward calculation we obtain,
\be\label{diff4}
\begin{split}
z^i(h)&-\hat z^i(h)=
\big(\bar F^i_{\;,j}\bar F^j_{\;,k}\bar F^k_{\;,l}
-\bar F^i_{\;,jk} \bar F^j \bar F^k_{\;,l}
\big)
w_5^l(h)\\
&+\frac{1}{2}\bar F^i_{\;,jk}\bar F^j_{\;,l}\bar F^k_{\;,m}
\bigg\{\int_0^h\diff t\,w_2^l(t)w_2^m(t)
+\frac{h^3}{180}w_1^l(h)w_1^m(h)
-\frac{h^2}{12}w_1^l(h)w_2^m(h)\\
&\qquad\qquad\qquad\quad~+\frac{h}{6}w_2^l(h)w_2^m(h)
+\frac{h}{6}w_1^l(h)w_3^m(h)
-w_2^l(h)w_3^m(h)
\bigg\}
+\mathcal{O}(h^{9/2})\;,
\end{split}
\ee
where
\be
w_5^i(h)=\s^i_{\;r}\int_0^h \diff t \bigg(-\frac{h^2t}{12}+\frac{ht^2}{4}-\frac{t^3}{6}\bigg)
\xi^r(t)\;.
\ee
The leading term written in the first line 
behaves as $\mathcal{O}(h^{7/2})$. Moreover, it vanishes on average, so that the average difference between the exact and approximate solutions comes only from the subleading terms of order $h^4$. The variance of this difference, on the other hand, comes from the square of the leading term and is of order $h^7$. Thus, we have proven the estimates (\ref{diff}) quoted in the main text. 
Note that the $\mathcal{O}(h^4)$ terms in (\ref{diff4}) 
are quadratic in the noise and are not eliminated by the condition (\ref{Hess}), implying that further improving the accuracy of the approximation would require dealing with non-Gaussian random variables, which is beyond the scope of our method.

We now state 
\textbf{\textit{Proposition 1:}} 
Consider the stochastic equation (\ref{NLstoch+}) in a finite time interval $0\leq t\leq T$. Let $z(t)$ be its solution for a given realization of the noise $\xi(t)$.
Let us split the interval into ${\cal N}$ steps of size $h$ and inside each step 
$t_n\leq t\leq t_{n+1}$, $t_n=nh$, 
substitute the equation by the ODE (\ref{Langevin2}) with the variables $\zeta^r_{a|n}$, $a=1,2,3$, defined as in eqs.~(\ref{newrandvars}) but with the integrals taken from $t_n$ to $t_{n+1}$.
Denote by $\hat z(t)$ the continuous solution of the obtained sequence of ODEs, such that 
$\hat z(0)=z(0)$. Then for small enough $h$ the difference between $z(t)$ and $\hat z(t)$, averaged over the realizations of the noise, satisfies the inequality (\ref{major3}) with some constant $C$ which does not depend on $h$.

We prove this proposition under a technical assumption that the first and second derivatives of the force $F^i(z)$ are bounded in the domain probed by the stochastic trajectories. It is likely that a proper refinement of the proof will be able to relax this assumption.

The idea is to relate the difference between the exact and the approximate solutions at the end and beginning of the $n$th time step. We write, 
\begin{align}\label{incr1}
\Delta z^i_{n+1}\equiv
z^i(t_{n+1})-\hat z^i(t_{n+1})=z^i(t_{n+1})-\breve{ z}^i(t_{n+1})+\breve{z}^i(t_{n+1})-\hat z^i(t_{n+1})\;,
\end{align}
where $\breve z(t)$ is the solution of the approximating ODE at $t_n\leq t\leq t_{n+1}$ satisfying the condition $\breve z(t_n)=z(t_n)$.
The previous discussion implies, 
\be
\label{stepdiff}
\big\langle \big(z(t_{n+1})-\breve z(t_{n+1})\big)^2\big\rangle
\leq C_1\,h^7\;,
\ee
where $C_1$ depends on the derivatives of the deterministic force and is bounded according to our assumption. Both $\breve z(t)$ and $\hat z(t)$ are solutions to the same regular ODE, with slightly different initial conditions at $t_n$. If these solutions are initially sufficiently close to each other, they cannot diverge much over a small time step. More in detail, the evolution of the difference between the two solutions 
is described by the equation
\be
\frac{d}{dt} (\breve z^i-\hat z^i)
=F^i_{\;,j}(\hat z) \,(\breve z^j-\hat z^j)
+F^i_{\;,jl}(\hat z) F^l_{\;,k}(\hat z) \s^k_{\;r} \zeta^r_{3|n}
\,(\breve z^j-\hat z^j)
+\mathcal{O}\big((\breve z-\hat z)^2\big)\;,
\ee
implying an estimate
\be
\label{ODEdiff}
|\breve z(t_{n+1})-\hat z(t_{n+1})|\leq (1+C_2\, h)\,|\breve z(t_n)-\hat z(t_n)|
\ee
valid for sufficiently small $|\breve z(t_n)-\hat z(t_n)|$. We now recall that $\breve z(t_n)$ coincides with the exact solution $z(t_n)$ and by combining eqs.~(\ref{incr1}), (\ref{stepdiff}), (\ref{ODEdiff}) obtain,
\be
\langle (\Delta z_{n+1})^2\rangle\leq (1+C_3 h)\, \langle (\Delta
z_{n})^2\rangle +C_1 h^7\;,
\ee
where we have used the fact the ODE solution at $t>t_n$ and the  
value $\Delta z_n$ are uncorrelated. 
Together with the initial condition $\Delta z_0=0$ 
this inequality leads to a bound
\be
\langle (\Delta z_n)^2\rangle\leq \frac{C_1h^6}{C_3}
\big[(1+C_3 h)^n-1\big]\;.
\ee
In the limit of small $h$ and fixed $nh=t_n$, the above expression becomes
\be
\langle (\Delta z_n)^2\rangle\leq \frac{C_1h^6}{C_3}
\big[\e^{C_3 t_n}-1\big] 
\leq \frac{C_1h^6}{C_3}
\big[\e^{C_3 T}-1\big] 
\;.
\ee
Since $\langle |\Delta z_n|\rangle \leq \sqrt{\langle (\Delta z_n)^2\rangle}$, we arrive at the bound (\ref{major3}) from the main text.

\section{Numerical algorithm}
\label{app:algorithm}

Below we present two exemplary algorithms to solve the Langevin \cref{Langevin0} using the PQ (\cref{PQ}) and LN (\cref{LN}) splittings with strong order $\gamma=3$ and splitting order $s=4$. For the LN splitting we assume that the variables $q^i$ diagonalize the linear part of the deterministic force: $f^i_{\rm lin}(q) = -\Omega_i^2 q^i$ (no summation).
The splitting coefficients are \cite{mclachlan1992accuracy}:
\be
\begin{aligned}
& a_1 = 0.5153528374311229364 \;, ~~ a_2 = -0.085782019412973646 \;, \\ 
& a_3 = 0.4415830236164665242\;, ~~ a_4 = 0.1288461583653841854 \;, \\
& b_1 = 0.1344961992774310892 \;, ~~ b_2 = -0.2248198030794208058 \;, \\ 
& b_3 = 0.7563200005156682911\;, ~~ b_4 = 0.3340036032863214255 \;.
\end{aligned}
\ee

\textbf{Algorithm 1: (PQ),  $(\gamma,s)=(3,4)$}
\hrule
\vspace{10pt}
Given $q^i=q^i_0$, $p^i=p^i_0$, evaluate $f^i(q_0)$, $i=1,...,d$

\textbf{for} $k=0,...,\mathcal{N}-1$ \textbf{do}

~~~~~sample $\zeta_a^r$, $a=1,2,3$, $r=1,...,R$, according to \cref{Cab}

~~~~~evaluate $\mu_1^i$, $\mu_2^i$ according to \cref{Mu12}

~~~~~\textbf{for} $l=1,...,s$ \textbf{do}

~~~~~~~~~~evaluate $v^i=f^i(q)+\mu_2^i$

~~~~~~~~~~update $p^i \leftarrow \frac{v^i}{\eta} + \e^{-b_l h \eta} \bigl( p^i - \frac{t^i}{\eta} \bigr)$

~~~~~~~~~~update $q^i \leftarrow q^i + a_l h (p^i+\mu_1^i)$

~~~~~\textbf{end for}

\textbf{end for}

\textbf{return} $q^i,p^i$

\vspace{10pt}
\textbf{Algorithm 2: (LN),  $(\gamma,s)=(3,4)$}
\hrule
\vspace{10pt}
Given $q^i=q^i_0$, $p^i=p^i_0$, evaluate $f^i(q_0)$, $i=1,...,d$

evaluate $\Omega_{i,\eta}=\sqrt{| \O_i^2-\eta^2/4 | }$

\textbf{for} $k=0,...,\mathcal{N}-1$ \textbf{do}

~~~~~sample $\zeta_a^r$, $a=1,2,3$, $r=1,...,R$, according to \cref{Cab}

~~~~~evaluate $\mu_1^i$, $\mu_2^i$ according to \cref{Mu12}

~~~~~\textbf{for} $l=1,...,s$ \textbf{do}

~~~~~~~~~~update $p^i \leftarrow p^i - b_l h \Omega_i^2 q^i$

~~~~~~~~~~evaluate $A_q^i = q^i - \frac{\eta\mu_1^i+ \mu_2^i}{\O_i^2}$

~~~~~~~~~~~~~~~~~~~~~ $B_q^i = \frac{1}{\O_{i, \eta}}\left(p^i + \frac{\eta q^i}{2} \right) + \frac{1}{\O_{i, \eta}}\left(\mu_1^i - \frac{\eta (\eta\mu_1^i+ \mu_2^i)}{2\O_i^2} \right) $

~~~~~~~~~~~~~~~~~~~~~ $C_q^i = \frac{\eta\mu_1^i+ \mu_2^i}{\O_i^2} $

~~~~~~~~~~~~~~~~~~~~~ $A_p^i = p^i + \mu_1^i $

~~~~~~~~~~~~~~~~~~~~~ $B_p^i = -\frac{1}{\O_{i,\eta}}\left( \O_i^2 q^i + \frac{\eta p^i}{2} \right) + \frac{\eta\mu_1^i+2\mu_2^i}{2\O_{i,\eta}} $

~~~~~~~~~~~~~~~~~~~~~ $C_p^i = -\mu_1^i $

~~~~~~~~~~\textbf{if} $\Omega_i<\eta/2$ \textbf{then}

~~~~~~~~~~~~~~~update $q^i \leftarrow \bigl( A_q^i \cos(a_l h \Omega_{i,\eta} ) + B_q^i\sin(a_l h \Omega_{i,\eta})    \bigr)\:\e^{-\eta a_l h/2}+ C_q^i  $

~~~~~~~~~~~~~~~update $p^i \leftarrow \bigl( A_p^i \cos(a_l h \Omega_{i,\eta} ) + B_p^i\sin(a_l h \Omega_{i,\eta})    \bigr)\:\e^{-\eta a_l h/2}+ C_p^i  $

~~~~~~~~~~\textbf{else}

~~~~~~~~~~~~~~~update $q^i \leftarrow \bigl( A_q^i \cosh(a_l h \Omega_{i,\eta} ) + B_q^i\sinh(a_l h \Omega_{i,\eta})    \bigr)\:\e^{-\eta a_l h/2}+ C_q^i  $

~~~~~~~~~~~~~~~update $p^i \leftarrow \bigl( A_p^i \cosh(a_l h \Omega_{i,\eta} ) + B_p^i\sinh(a_l h \Omega_{i,\eta})    \bigr)\:\e^{-\eta a_l h/2}+ C_p^i  $

~~~~~~~~~~\textbf{end if}

~~~~~\textbf{end for}

\textbf{end for}

\textbf{return} $q^i,p^i$

\section{Linear evolution for the pseudo-spectral scheme}
\label{app:pseudo} 

Here we give explicit expressions for the solution of eq.~(\ref{EqsL2}) describing the linear evolution of the field Fourier modes in the pseudo-spectral scheme of Sec.~\ref{ssec:fieldtheory}. Since the modes evolve independently, we can focus on a single mode 
with frequency $\Omega$ and omit the index $j$ labeling the modes.
Let us   
denote the amplitude of the mode and the corresponding momentum at time $t$ by $\tilde \phi_t$ and $\tilde \pi_t$. If the mode is under-damped, $\Omega>\eta/2$, 
the solution at time $t+ha$ takes the form,
\be \label{SolL2}
\e^{ha\hat{A}}: \quad \begin{array}{l}
\tilde\phi_{t+ha} = \big( A_{\phi} \cos(ha\O_{\eta}) + B_{\phi}\sin(ha\O_{\eta})\big) \e^{-\frac{\eta ha}{2}} + C_{\phi} \;, \\
\tilde\pi_{t+ha} = \big( A_{\pi} \cos(ha\O_{\eta}) + B_{\pi}\sin(ha\O_{\eta})\big)\e^{-\frac{\eta ha}{2}} + C_{\pi} \;.
\end{array}
\ee
Here $\O_{\eta} = \sqrt{\O^2-\eta^2/4}$
and
\bseq
\begin{align}
& A_{\phi} = \tilde\phi_t - \frac{\eta\tilde\mu_1+ \tilde\mu_2}{\O^2} \;, \\
& B_{\phi} = \frac{1}{\O_{\eta}}\left(\tilde \pi_t + \frac{\eta \tilde\phi_t}{2} \right) + \frac{1}{\O_{\eta}}\left( \tilde\mu_1 - \frac{\eta (\eta\tilde\mu_1+ \tilde\mu_2)}{2\O^2} \right) \;, \\
& C_{\phi} = \frac{\eta\tilde\mu_1+ \tilde\mu_2}{\O^2} \;, \\
& A_{\pi} = \tilde\pi_t + \tilde\mu_1 \;, \\
& B_{\pi} = -\frac{1}{\O_{\eta}}\left( \O^2 \tilde\phi_t + \frac{\eta\tilde \pi_t}{2} \right) + \frac{\eta\tilde\mu_1+2\tilde\mu_2}{2\O_{\eta}} \;, \\
& C_{\pi} = -\tilde\mu_1  \;.
\end{align}
\eseq
For the over-damped modes, $\O<\eta/2$, we take $\O_{\eta} = \sqrt{\eta^2/4-\O^2}$ and  replace $\cos, \sin\mapsto\cosh, \sinh$ in the above expressions.

\section{Precision of the strong convergence tests}
\label{app:strong}

\subsection{Evaluation of stochastic integrals}

To measure the rate of strong convergence, we need to compute the stochastic integrals (\ref{newrandvars}) for a given realization of the white noise. Naively, one could think that it is sufficient to generate a realization of the Wiener process $W(t)$ on a dense grid $t_k=k\ve$ with 
$\ve \ll h$ and then evaluate (\ref{newrandvars}) using the formulas,
\bseq
\label{xiWien}
\begin{align}
\label{xiWien1}
&\zeta_1=\frac{1}{h}\big(W(h)-W(0)\big)\;,\\
\label{xiWien2}
&\zeta_2=-\frac{1}{2}\big(W(h)+W(0)\big)+\frac{1}{h}\int_0^h \diff t W(t)\;,\\
\label{xiWien3}
&\zeta_3=\frac{h}{12}\big(W(h)-W(0)\big)+
\frac{1}{h}\int_0^h \diff t \left(\frac{h}{2}-t\right) W(t)\;.
\end{align}
\eseq
The first expression (\ref{xiWien1}) is exact as long as $h$
is an integer multiple of $\ve$, 
\be\label{heps}
h=K \ve\;.
\ee
A problem arises, however,
with the integrals appearing in eqs.~(\ref{xiWien2}) and (\ref{xiWien3}). Since the Wiener process is not differentiable, the root-mean-square error between the discrete and continuous versions of these integrals will always scale as the first power of $\ve$, no matter what discretization method we use. 

In more detail, the discretization error estimate for the integral in eq.~(\ref{xiWien2}) is $A\cdot\ve/\sqrt{h}$, where $A$ is a numerical coefficient depending on the chosen discretization scheme.   
For the convergence tests, this
error must be smaller than the error of the stochastic approximation
eq.~(\ref{Langevin2+}). This implies that to test the 2nd and 3rd order convergence we
need, respectively, 
\bseq
\begin{align}
\label{req2}
\ve/\sqrt{h}\ll h^{3/2}~~~ \Longrightarrow ~~~ \ve\ll h^2\;,\\
\label{req3}
\ve/\sqrt{h}\ll h^{5/2}~~~ \Longrightarrow ~~~ \ve\ll h^3\;.
\end{align}
\eseq
These conditions are prohibitively 
stringent. For example, a test of the 3rd order scheme down to $h\sim 10^{-3}$ would require
extremely small grid step $\ve\ll 10^{-9}$.

The above problem is resolved by the same approach which led
us to \cref{Langevin2}. Instead of a single Wiener process, we need
to generate on the grid several processes which encode also
the higher moments of the noise. Consider three
discrete random
processes $W_k^{(1)}$, $W_k^{(2)}$, $W_k^{(3)}$, such that 
$\Delta W_k^{(a)}\equiv W_{k+1}^{(a)}-W_k^{(a)}$ are independent Gaussian random
variables with the correlation matrix (cf. \cref{Cab}),
\bseq
\label{Wcorr}
\begin{align}
&\big\langle \Delta W^{(a)}_{k}\Delta W_{k'}^{(b)}\big\rangle ={\cal C}_{\ve}^{(a)}\,\delta^{ab}\,\delta_{kk'}\;,\\
&\mathcal{C}_\ve^{(1)}=\ve\;,~~~~
\mathcal{C}_\ve^{(2)}=\frac{\ve^3}{12}\;,~~~~
\mathcal{C}_\ve^{(3)}=\frac{\ve^5}{720}\;.
\end{align}
\eseq
They describe realizations of the following integrals:
\bseq
\label{newWiens}
\begin{align}
\label{newWien0}
&\Delta W_k^{(1)}=\int_{t_k}^{t_{k+1}} \diff t\,\xi(t)\;,\\
\label{newWien1}
&\Delta
W_k^{(2)}=\int_{t_k}^{t_{k+1}}\diff t\bigg(\frac{\ve}{2}-(t-t_k)\bigg)\xi(t)\;,\\
\label{newWien2}
&\Delta
W_k^{(3)}=\int_{t_k}^{t_{k+1}}\diff t\bigg(\frac{\ve^2}{12}-\frac{\ve}{2}(t-t_k)
+\frac{(t-t_k)^2}{2}\bigg)\xi(t)\;.
\end{align}
\eseq
Then under the condition (\ref{heps}) the {\em exact} expressions for the noise variables
(\ref{newrandvars}) read,
\bseq
\label{xiWienNew}
\begin{align}
\label{xiWienNew1}
&\zeta_1=\frac{1}{h}\sum_{k=0}^{K-1}\Delta W_k^{(1)}\;,\\
\label{xiWienNew2}
&\zeta_2=\frac{1}{h}\sum_{k=0}^{K-1}\bigg[\Delta W_k^{(2)}+
\bigg(\frac{h}{2}-t_k-\frac{\ve}{2}\bigg)\Delta W_k^{(1)}\bigg]\;,\\
\label{xiWienNew3}
&\zeta_3=\frac{1}{h}\sum_{k=0}^{K-1}\bigg[\Delta W_k^{(3)}+
\bigg(\frac{h}{2}-t_k-\frac{\ve}{2}\bigg)\Delta W_k^{(2)}
+\bigg(\frac{h^2}{12}-\frac{ht_k}{2}+\frac{t_k^2}{2}-\frac{\ve h}{4}
+\frac{\ve t_k}{2}+\frac{\ve^2}{6}
\bigg)\Delta W_k^{(1)}
\bigg].
\end{align}
\eseq
These formulas allow us to take $\ve=h_*$ where $h_*$ is the minimal time step used in the convergence tests.
We use this method in Sec.~\ref{sec:particle} to measure the strong order of different stochastic schemes.

\subsection{Round-off errors}

When testing the strong convergence, one needs to make sure that the result is not spoiled by the round-off errors due to the finite machine precision. To estimate when this can become an issue, let us assume that the round-off error accumulates linearly with the total number of time steps made in the computation. Then the total discrepancy between the numerical solution and its continuum limit will be 
\be \label{Diff1}
\langle |z_h(T) - z(T)|\rangle \simeq C h^\g + \frac{T}{h}\epsilon \;,
\ee 
where $\g$ is the strong convergence order of the scheme, $C$ is an order-one coefficient, $T$ is the total evolution time, and $\epsilon$ is the round-off error at each step. The discrepancy is minimized at the optimal time step $h_{\rm opt}\simeq (T\epsilon)^{1/\g}$; at smaller steps, the discrepancy is dominated by the round-off error. For $\g=3$, $T=100$ and $\epsilon=10^{-16}$, which corresponds to the 64-bit machine precision type `double', we obtain $h_{\rm opt}\simeq 3\cdot 10^{-4}$. The minimal value of the discrepancy is then 
$\langle |\Delta z_T|\rangle_{\rm min}\simeq 3\cdot 10^{-11}$.

In our tests in Sec.~\ref{sec:particle} we cannot compare the numerical solution to its continuum limit since we do not know the latter. Instead, we compare the numerical solutions found using different time steps. If the reference step $h_*$ is smaller than $h_{\rm opt}$, the measured discrepancy is estimated as 
\be \label{Diff2}
\langle |\Delta z_T|\rangle\equiv 
\langle |z_{h}(T) - z_{h_*}(T)|\rangle 
\simeq Ch^\g + \frac{T\epsilon}{h_*} \;.
\ee 
The second term becomes dominant at $h$ below
\be
\label{hmin}
h_{\rm min}\simeq (T\epsilon/h_*)^{1/\g}\;,
\ee
which for our fiducial values $\g=3$, $T=100$, $h_*=2\cdot 10^{-5}$ and $\epsilon\simeq 10^{-16}$ gives $h_{\rm min}\simeq 8\cdot10^{-4}$. This is quite large and does not allow us to reliably assess the convergence of the scheme below $\langle |\Delta z_T|\rangle\sim 10^{-9}$. To avoid this problem, we use the 128-bit machine precision type `float\underline{ }128'
with $\epsilon\simeq 10^{-34}$.

\begin{figure}[t]
    \begin{center}
    \begin{minipage}[h]{0.55\linewidth}
        \centering
        \includegraphics[width=1.0\linewidth]{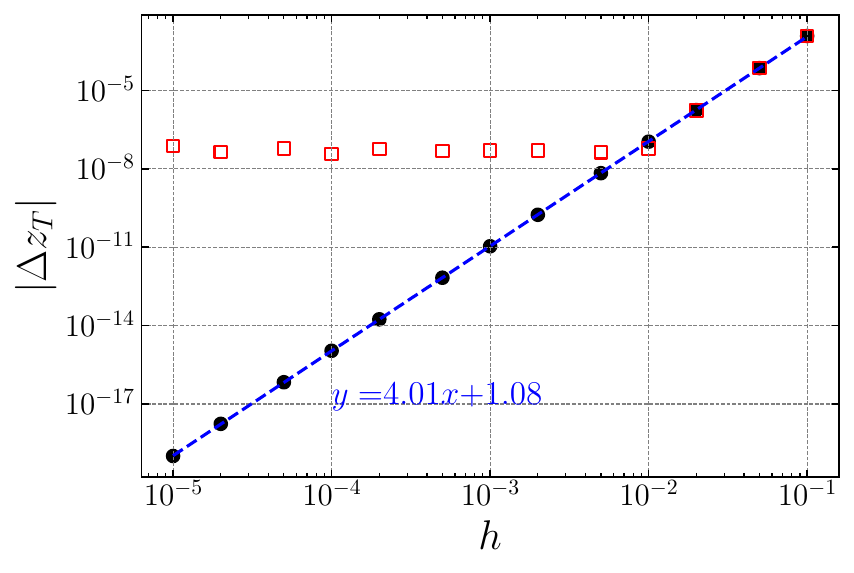}
	\end{minipage}
    \end{center}
    \caption{Difference between the numerical solutions to the deterministic pendulum equation obtained with the time step $h$ and the reference solution with $h_*=5\cdot10^{-6}$. We use the 4th order operator-splitting method with PQ splitting. Red squares (black dots) show the results obtained using the machine precision `double' (`float\underline{ }128'). The straight line shows a power-law fit to the data points. 
    }
	\label{fig:pend_Ham}
\end{figure}

We have confirmed the estimate (\ref{Diff2}) with an explicit computation. Since the estimate does not depend on whether we are solving stochastic or deterministic equation, we performed the check for the pendulum equation with the force (\ref{Pot_pend}) and vanishing friction and noise ($\eta=\s=0$). We solved it using the 4th order splitting method and measured the discrepancy between numerical solutions with different steps $h$ and the reference solution with step ${h_*=5\cdot10^{-6}}$. The results obtained using the 64-bit arithmetic are shown by red squares in Fig.~\ref{fig:pend_Ham}. We see that they hit the floor $|\Delta z_T|\sim 5\cdot 10^{-8}$ at $h\lesssim 10^{-2}$. This is consistent with the estimate (\ref{hmin}), where we set $\g=4$. We conclude that the round-off error indeed accumulates linearly with the number of steps. Figure~\ref{fig:pend_Ham} also shows the results of the same computation done with the 128-bit arithmetic (black dots). Clearly, the round-off errors are negligible in this case.

\section{Thermal spectrum of $\l\phi^4$ theory in 2d}
\label{app:loops}

The conservative part of eq.~(\ref{LangEq}) can be obtained from the Hamiltonian,
\be 
\label{S1}
    H = \int\diff x\left(\frac{\pi^2}{2}+\frac{\phi'^2}{2}
    +\frac{\phi^2}{2} + \frac{\a \phi^4}{4}\right)\;.
\ee
This gives rise to the (classical) partition function of the system in the form of path integral, 
\be
\label{partit}
Z=\int [d\phi] \exp\bigg\{-\frac{1}{\T}\int \diff x\bigg(\frac{\phi'^2}{2}
    +\frac{\phi^2}{2} + \frac{\a \phi^4}{4}\bigg)\bigg\}\;,
\ee
where we have integrated out the canonical momentum $\pi(x)$. Perturbative expansion of eq.~(\ref{partit}) in the powers of the interaction term $\phi^4$ gives rise to the standard diagrammatic technique for the thermal correlators of the field $\phi$, with the propagator and the vertex: 
\be
\label{propag}
\begin{fmffile}{philine}
\parbox{60pt}{\begin{fmfgraph*}(40,40)
\fmfpen{thick}
\fmfleft{i} 
\fmfright{o}
\fmf{plain,label=$k$,label.side=left}{i,o}
\end{fmfgraph*}}
\end{fmffile}
=\frac{\T}{k^2+1}\;,~~~~~~~~~~~~
\begin{fmffile}{phivert}
\parbox{60pt}{\begin{fmfgraph*}(50,30)
\fmfpen{thick}
\fmfleft{i1,i2} 
\fmfright{o1,o2}
\fmfdot{v}
\fmf{plain}{i1,v}
\fmf{plain}{i2,v}
\fmf{plain}{o1,v}
\fmf{plain}{o2,v}
\end{fmfgraph*}}
\end{fmffile}
=-\frac{\a}{4\T}\;.
\ee
The expansion parameter is identified with the temperature $\T$. 

Defining the power spectrum in the Fourier space,
\be
\big\langle \tilde \phi(k)\tilde\phi(k')\big\rangle=
(2\pi)\delta(k+k')\,\PP(k)\;,
\ee
we express it through the polarization operator,
\be
\label{PPi}
\PP(k)=\T\big(k^2+1+\Pi(k^2)\big)^{-1}\;,
\ee
where the latter is given by the sum of 1-particle irreducible loop diagrams. Up to 3 loops we have: 
\[
\begin{split}
\Pi(k^2)&=-\T\Bigg[~~
\begin{fmffile}{Pi1}
\parbox{50pt}{\begin{fmfgraph*}(40,50)
\fmfpen{thick}
\fmfleft{i} 
\fmfright{o}
\fmfdot{v}
\fmfv{l=$~\Pi_{1}$,l.a=-90,l.d=8mm}{v}
\fmf{plain}{i,v}
\fmf{plain}{o,v}
\fmf{plain,tension=1/2}{v,v}
\end{fmfgraph*}}
\end{fmffile}
+~~
\begin{fmffile}{Pi21}
\parbox{50pt}{\begin{fmfgraph*}(40,100)
\fmfpen{thick}
\fmfleft{i2,i4,i,i1,i3} 
\fmfright{o2,o4,o,o1,o3}
\fmfdot{v}
\fmf{plain}{i,v,o}
\fmf{plain,tension=0.55}{v,v}
\fmffreeze
\fmf{phantom}{i1,v1,o1}
\fmfdot{v1}
\fmf{plain,tension=0.5}{v1,v1}
\fmfv{l=$~~\Pi_{2,,1}$,l.a=-90,l.d=8mm}{v}
\end{fmfgraph*}}
\end{fmffile}
+~~
\begin{fmffile}{Pi22}
\parbox{70pt}{\begin{fmfgraph*}(60,40)
\fmfpen{thick}
\fmfleft{i}
\fmfright{o}
\fmf{plain,tension=2}{i,v1}
\fmf{plain,tension=2}{o,v2}
\fmf{plain,tension=1/3}{v1,v3,v2}
\fmf{plain,left,tension=1/3}{v1,v2,v1}
\fmfdot{v1,v2}
\fmfv{l=$~~\Pi_{2,,2}$,l.a=-90,l.d=8mm}{v3}
\end{fmfgraph*}}
\end{fmffile}\\
&+~~
\begin{fmffile}{Pi31}
\parbox{50pt}{\begin{fmfgraph*}(40,120)
\fmfpen{thick}
\fmfleft{i2,i4,i6,i,i1,i3,i5} 
\fmfright{o2,o4,o6,o,o1,o3,o5}
\fmfdot{v}
\fmf{plain}{i,v,o}
\fmf{plain,tension=0.65}{v,v}
\fmffreeze
\fmf{phantom}{i1,v1,o1}
\fmfdot{v1}
\fmf{plain,tension=0.65}{v1,v1}
\fmffreeze
\fmf{phantom}{i3,v3,o3}
\fmfdot{v3}
\fmf{plain,tension=0.6}{v3,v3}
\fmfv{l=$~~\Pi_{3,,1}$,l.a=-90,l.d=8mm}{v}
\end{fmfgraph*}}
\end{fmffile}
+~~
\begin{fmffile}{Pi32}
\parbox{50pt}{\begin{fmfgraph*}(40,100)
\fmfpen{thick}
\fmfleft{i2,i4,i6,i,i1,i3,i5} 
\fmfright{o2,o4,o6,o,o1,o3,o5}
\fmfdot{v}
\fmf{plain}{i,v,o}
\fmf{plain,tension=0.42}{v,v}
\fmffreeze
\fmf{phantom}{i5,v1,v2,o1}
\fmf{plain,right,tension=0.5}{v2,v2}
\fmf{phantom}{o5,v3,v4,i1}
\fmf{plain,right=90,tension=0.5}{v4,v4}
\fmfdot{v2,v4}
\fmfv{l=$~~\Pi_{3,,2}$,l.a=-90,l.d=8mm}{v}
\end{fmfgraph*}}
\end{fmffile}
+~~
\begin{fmffile}{Pi33}
\parbox{70pt}{\begin{fmfgraph*}(60,100)
\fmfpen{thick}
\fmfleft{i}
\fmfright{o}
\fmf{plain,tension=2}{i,v1}
\fmf{plain,tension=2}{o,v2}
\fmf{plain,tension=1/3}{v1,v3,v2}
\fmf{plain,left,tension=1/3}{v1,v2,v1}
\fmfdot{v1,v2}
\fmffreeze
\fmfforce{(.5w,.67h)}{v4}
\fmfdot{v4}
\fmf{phantom}{i,v4,o}
\fmf{plain,tension=1.1}{v4,v4}
\fmfv{l=$~~\Pi_{3,,3}$,l.a=-90,l.d=8mm}{v3}
\end{fmfgraph*}}
\end{fmffile}
+~~
\begin{fmffile}{Pi34}
\parbox{70pt}{\begin{fmfgraph*}(60,100)
\fmfpen{thick}
\fmfleft{i6,i4,i2,i,i1,i3,i5}
\fmfright{o6,o4,o2,o,o1,o3,o5}
\fmf{plain}{i,v,o}
\fmfdot{v}
\fmffreeze
\fmf{phantom,tension=5}{i3,v1}
\fmf{phantom,tension=5}{o3,v2}
\fmf{plain,tension=1/3}{v1,v2}
\fmf{plain,left=0.5,tension=1/3}{v1,v2,v1}
\fmfdot{v1,v2}
\fmffreeze
\fmf{plain,right=0.1}{v1,v}
\fmf{plain,left=0.1}{v2,v}
\fmfv{l=$~~\Pi_{3,,4}$,l.a=-90,l.d=8mm}{v}
\end{fmfgraph*}}
\end{fmffile}
+~~
\begin{fmffile}{Pi35}
\parbox{70pt}{\begin{fmfgraph*}(60,100)
\fmfpen{thick}
\fmfleft{i6,i4,i2,i,i1,i3,i5}
\fmfright{o6,o4,o2,o,o1,o3,o5}
\fmf{plain,tension=1}{i,v1}
\fmf{plain,tension=1}{o,v2}
\fmf{plain,tension=2/3}{v1,v3,v2}
\fmfdot{v1,v2}
\fmffreeze
\fmf{phantom}{i3,v4,o3}
\fmfdot{v4}
\fmffreeze
\fmf{plain,right=0.5}{v4,v1}
\fmf{plain,left=0.5}{v4,v2}
\fmf{plain,left=0.2}{v4,v1}
\fmf{plain,right=0.2}{v4,v2}
\fmfv{l=$~~\Pi_{3,,5}$,l.a=-90,l.d=8mm}{v3}
\end{fmfgraph*}}
\end{fmffile}
\Bigg]
\end{split}
\]
The respective contributions are all finite and read:
\bseq
\begin{align}
\Pi_1=&-12\cdot\left(-\frac{\a\T}{4}\right)\cdot\int \frac{dq}{(2\pi)(q^2+1)}=\frac{3\a\T}{2}\;,\\
\Pi_{2,1}=&-288\cdot\frac{1}{2!}\left(-\frac{\a\T}{4}\right)^2\cdot
\int\frac{dq_1}{(2\pi)(q_1^2+1)^2}\int\frac{dq_2}{(2\pi)(q_2^2+1)}=-\frac{9\a^2\T^2}{8}\;,\\
\Pi_{2,2}=&-192\cdot\frac{1}{2!}\left(-\frac{\a\T}{4}\right)^2\cdot
\int\frac{dq_1dq_2}{(2\pi)^2(q_1^2+1)(q_2^2+1)((k+q_1+q_2)^2+1)}=-\frac{9\a^2\T^2}{2(k^2+9)}\;,\\
\Pi_{3,1}=&-10368\cdot\frac{1}{3!}\left(-\frac{\a\T}{4}\right)^3\cdot
\left(\int\frac{dq_1}{(2\pi)(q_1^2+1)^2}\right)^2 \int \frac{dq_2}{(2\pi)(q_2^2+1)}=
\frac{27\a^3\T^3}{32}\;,\\
\Pi_{3,2}=&-10368\cdot\frac{1}{3!}\left(-\frac{\a\T}{4}\right)^3\cdot
\int\frac{dq_1}{(2\pi)(q_1^2+1)^3}\left(\int \frac{dq_2}{(2\pi)(q_2^2+1)}\right)^2=
\frac{81\a^3\T^3}{64}\;,\\
\Pi_{3,3}=&-20736\cdot\frac{1}{3!}\left(-\frac{\a\T}{4}\right)^3\cdot
\int\frac{dq_1dq_2}{(2\pi)^2(q_1^2+1)(q_2^2+1)((k+q_1+q_2)^2+1)^2}\notag\\
&\qquad\qquad\qquad\qquad\qquad\qquad
\times
\int\frac{dq_3}{(2\pi)(q_3^2+1)}
=\frac{27(k^2+18)}{4(k^2+9)^2}\a^3\T^3\;,\\
\Pi_{3,4}=&-6912\cdot\frac{1}{3!}\left(-\frac{\a\T}{4}\right)^3\cdot
\int\frac{dq_1dq_2dq_3}{(2\pi)^3(q_1^2+1)(q_2^2+1)(q_3^2+1)}\notag\\
&\qquad\qquad\qquad\qquad\qquad\qquad
\times\frac{1}{((q_1+q_2+q_3)^2+1)^2}=\frac{45\a^3\T^3}{128}\;,\\
\Pi_{3,5}=&-20736\cdot\frac{1}{3!}\left(-\frac{\a\T}{4}\right)^3\cdot
\int\frac{dq_1dq_2dq_3}{(2\pi)^3(q_1^2+1)(q_2^2+1)(q_3^2+1)}\notag\\
&\qquad\qquad\qquad\qquad
\times\frac{1}{((k+q_1+q_2)^2+1)((q_3-q_1-q_2)^2+1)}
=\frac{27(k^2+45)}{16(k^2+9)^2}\a^3\T^3\;.
\end{align}
\eseq
Combining these terms together and substituting into eq.~(\ref{PPi}) we obtain eq.~(\ref{psTh}) from the main text.

\bibliographystyle{JHEP}
\bibliography{Refs}

\end{document}